\documentclass[structabstract, longauth]{aa}  
\usepackage{graphicx}
\usepackage{txfonts}

\usepackage{natbib}
\bibpunct[, ]{(}{)}{;}{a}{}{,}
\newcommand{\griz}{$g^\prime r^\prime i^\prime z^\prime$~}

\newcommand{\gK}{$g^\prime r^\prime i^\prime z^\prime JHK_s$~}
\newcommand{\Msun}{\ensuremath{M_{\odot}}}

\begin{document}
\title{The SEDs and Host Galaxies of the dustiest GRB afterglows\thanks{Based on observations made with GROND at the MPI/ESO 2.2~m telescope and with telescopes at the European Southern Observatory at LaSilla/Paranal, Chile under program 086.A-0533 and obtained from the ESO/ST-ECF Science Archive Facility from programs 177.A-0591 and 078.D-0416.}}

\author{T.~Kr\"{u}hler \inst{1,2, 3}
          \and
          J.~Greiner \inst{1}
          \and 
          P.~Schady \inst{1}
          \and
          S.~Savaglio \inst{1}
          \and 
          P.~M.~J.~Afonso \inst{1,4} 
          \and
          C.~Clemens \inst{1}
          \and
          J.~Elliott \inst{1}
          \and
          R.~Filgas \inst{1}
          \and
          D.~Gruber \inst{1} 
          \and
          D.~A. Kann \inst{5}       
          \and
          S.~Klose \inst{5}
          \and 
          A.~ K\"{u}pc\"{u}-Yolda\c{s} \inst{6}
          \and 
          S.~McBreen \inst{7}
          \and
          F.~Olivares E. \inst{1} 
          \and
          D.~Pierini \thanks{Visiting Astronomer at MPE}
          \and
          A.~Rau \inst{1}
          \and 
          A.~Rossi \inst{5}
          \and    
		  M.~Nardini \inst{1,8}
		  \and 
		  A.~Nicuesa Guelbenzu \inst{5}
		  \and
		  V.~Sudilovsky \inst{1}
		  \and
		  A.~C.~Updike \inst{9, 10}
          }


\institute{Max-Planck-Institut f\"{u}r extraterrestrische Physik, Giessenbachstra\ss e, 85748 Garching, Germany.
            \email{kruehler@mpe.mpg.de}
            \and
            Excellence Cluster Universe, Technische Universit\"{a}t M\"{u}nchen, Boltzmannstra\ss e 2, 85748, Garching, Germany.
            	\and
	Dark Cosmology Centre, Niels Bohr Institute, University of Copenhagen, Juliane Maries Vej 30, 2100 Copenhagen, Denmark.
	\and
           American River College, Physics \& Astronomy Dpt., 4700 College Oak Drive, Sacramento, CA 95841, USA.
           \and
           Th\"{u}ringer Landessternwarte Tautenburg, Sternwarte 5, 07778 Tautenburg, Germany.
           \and
           Institute of Astronomy, University of Cambridge, Madingley Road, CB3 0HA, Cambridge, UK.
           \and
           School of Physics, University College Dublin, Dublin 4, Ireland.
           \and
           Universit\`a degli studi di Milano-Bicocca, Piazza della Scienza 3, 20126 Milano, Italy.
           \and
           CRESST and the Observational Cosmology Laboratory, NASA/GSFC, Greenbelt, MD 20771, USA.
           \and
           Department of Astronomy, University of Maryland, College Park, MD 20742, USA.
           }


\date{} 

 
\abstract
  {The afterglows and host galaxies of long gamma-ray bursts (GRBs) offer unique opportunities to study star-forming galaxies in the high-$z$ Universe. Until recently, however, the information inferred from GRB follow-up observations was mostly limited to optically bright afterglows, biasing all demographic studies against sight-lines that contain large amounts of dust.}
{Here we present afterglow and host observations for a sample of bursts that are exemplary of previously missed ones because of high visual extinction ($A_V^{\rm GRB}\gtrsim1\,\rm{mag}$) along the sight-line. This facilitates an investigation of the properties, geometry and location of the absorbing dust of these poorly-explored host galaxies, and a comparison to hosts from optically-selected samples.}
{This work is based on GROND optical/NIR and \textit{Swift}/XRT X-ray observations of the afterglows, and multi-color imaging for eight GRB hosts. The afterglow and galaxy spectral energy distributions yield detailed insight into physical properties such as the dust and metal content along the GRB sight-line as well as galaxy-integrated characteristics like the host's stellar mass, luminosity, color-excess and star-formation rate.}
{For the eight afterglows considered in this study we report for the first time the redshift of GRB~081109 ($z=0.9787\pm0.0005$), and the visual extinction towards GRBs~081109 ($A_V^{\rm GRB}=3.4_{-0.3}^{+0.4}\,\rm{mag}$) and 100621A ($A_V^{\rm GRB}=3.8\pm0.2\,\rm{mag}$), which are among the largest ever derived for GRB afterglows. Combined with non-extinguished GRBs, there is a strong anti-correlation between the afterglow's metals-to-dust ratio and visual extinction. 

The hosts of the dustiest afterglows are diverse in their properties, but on average redder ($\langle(R-K)_{\rm AB} \rangle \sim 1.6\,\rm{mag}$), more luminous ($\langle L \rangle \sim 0.9\,L^{\ast}$) and massive ($\langle \log M_\ast [\Msun]\rangle \sim 9.8$) than the hosts of optically-bright events. We hence probe a different galaxy population, suggesting that previous host samples miss most of the massive, chemically-evolved and metal-rich members. This also indicates that the dust along the sight-line is often related to host properties, and thus probably located in the diffuse ISM or interstellar clouds and not in the immediate GRB environment. Some of the hosts in our sample, are blue, young or of small stellar mass illustrating that even apparently non-extinguished galaxies possess very dusty sight-lines due to a patchy dust distribution.}
{The afterglows and host galaxies of the dustiest GRBs provide evidence for a complex dust geometry in star-forming galaxies. In addition, they establish a population of luminous, massive and correspondingly chemically-evolved GRB hosts. This suggests that GRBs trace the global star-formation rate better than studies based on optically-selected host samples indicate, and the previously-claimed deficiency of high-mass host galaxies was at least partially a selection effect.}


\maketitle
%

\section{Introduction}

Long gamma-ray bursts (GRBs, see e.g., \citealt[][for reviews]{2007ChJAA...7....1Z, 2009ARA&A..47..567G}) are linked to core-collapse supernovae and hence star-formation via the death of massive stars \citep[e.g.,][]{1998Natur.395..670G, 2003Natur.423..847H}. At high redshifts, where a significant fraction of the star-formation is thought to be dust-obscured \citep[e.g.,][]{2000ApJ...544..218A, 2005ApJ...622..772C}, GRBs and their host galaxies offer an independent track towards a better understanding and full census of the star-formation in the early Universe: GRBs, having luminous emission in a simple power-law spectrum provide the ideal background light to illuminate dust-enshrouded star-forming regions which would otherwise remain unexplored, while at the same time pinpointing their host galaxies. 

However, the extent to which GRB hosts provide an unbiased picture of the formation of high-mass stars, and whether they preferentially occur in low-metallicity environments remains a much debated issue \citep[e.g.,][]{2003A&A...400..499L, 2003A&A...406L..63F, 2004MNRAS.352.1073T,2006Natur.441..463F, 2009ApJ...702..377K}. In single progenitor models, metal-poor stars are favored by theory \citep{1993ApJ...405..273W, 1999ApJ...524..262M}, as they would in principle be able to keep more angular momentum at the time of stellar collapse due to smaller wind pressures and losses throughout their evolution \citep[e.g.,][]{2005A&A...443..643Y, 2007A&A...473..603M}. However, binary progenitor channels could also play an important role in the formation of long GRBs \citep[e.g.,][]{1999ApJ...526..152F}, having somewhat relaxed metallicity constraints relative to single star progenitors \citep{2007PASP..119.1211F}. Observations of GRB hosts are hence not only important in a cosmological context, but provide relevant clues to the exact nature of GRB progenitors.

A fundamental limit of hitherto available GRB host galaxy samples is the incompleteness which arises from the non-detection of the optical afterglow of a GRB \citep[e.g.,][]{1998ApJ...493L..27G, 2001A&A...369..373F}. These optically dark bursts could be caused by either high-redshift \citep[e.g.,][]{2009ApJ...693.1610G, 2009Natur.461.1254T, 2009Natur.461.1258S, 0429BCucc}, large columns of dust \citep[e.g.,][]{2000ApJ...545..271K, 2003ApJ...592.1025K, 2008MNRAS.388.1743T, 2010arXiv1009.0004P} or an intrinsically fainter optical afterglow as compared to the extrapolation of X-ray data when using synchrotron emission theory, i.e., a decoupled optical/X-ray afterglow light-curve \citep[e.g.,][]{2006MNRAS.369.2059P, 2009MNRAS.393..253G, 2010MNRAS.403.1131N}. New afterglow samples became available through the recent advent of dedicated afterglow follow-up campaigns on medium-to-large aperture telescopes \citep[e.g.,][]{2009ApJS..185..526F, 2009ApJ...693.1484C, 2011A&A...526A..30G}. These new afterglow samples reach completeness levels of $\sim$90\% \citep{2011A&A...526A..30G} and settled the dark-burst issue: Around three quarters of dark bursts are the result of a dusty afterglow line of sight \citep[e.g.,][]{2009AJ....138.1690P, 2011A&A...526A..30G}. Accurate positions from afterglow observations are necessary to unambiguously associate galaxies to GRBs. The lack of optical/NIR afterglows for dark GRBs therefore implies an inherent bias against the associated host galaxies.

The available host population is not as complete as the most recent afterglow samples. Instead, it is largely selected from optically bright afterglows and shows a prevalence of young and vigorously star-forming galaxies with sub-$L^{\ast}$ luminosities and masses around $10^{9}$~\Msun~\citep[e.g.,][referenced as SGL09, hereafter]{1998ApJ...507L..25B, 2003A&A...400..499L, 2004A&A...425..913C, 2006Natur.441..463F, 2009ApJ...691..182S}. However, it is an open question whether this is a physical consequence of GRBs preferring low-metallicity environments, or merely a selection effect: Host galaxies of dark GRBs were typically not identified, and hence are under-represented in the available host sample.

Whether the physical characteristics of hosts of optically dark and bright GRBs are distinct is also the subject of discussion. Previous sample studies \citep[e.g.,][]{2003ApJ...588...99B, 2003A&A...400..499L, 2009AJ....138.1690P} have not revealed strong evidence of a significant difference. A handful of single dark GRBs were however hosted by red and dusty galaxies with high metallicities and stellar masses over $10^{11}$~\Msun~\citep[e.g.,][]{2006ApJ...647..471L, 2007ApJ...660..504B, 2010ApJ...712L..26L, 2010ApJ...719..378H, 2010A&A...515L...2K, 2010ApJ...723L.218C}. Recently \citet{2010AAS...21540509P, 2011AAS...21710802P} suggested, that the general galaxy population hosting dark bursts is redder and more luminous and hence suggestive of a higher metallicity than those selected via optically bright afterglows.
%


In this paper, we study the nature of GRB hosts that previously escaped detection due to the dust bias, and are hence exemplary of those missing from demographic studies. We avail of dedicated GRB afterglow campaigns with high completeness to preselect the GRB hosts for this study. These afterglow data do not only provide accurate positions for host identifications, but for the first time allow us to directly select dust extinguished (and not only optically faint) GRBs via well-sampled broad-band (NIR to X-ray) afterglow observations.

After selecting afterglows with visual extinctions $A_V^{\rm GRB}$ exceeding unity, we search for the associated hosts with the Gamma-Ray Optical and Near-Infrared Detector (GROND, \citealp{2008PASP..120..405G}) as well as the ESO Very Large and New Technology Telescopes (VLT and NTT, respectively) and the Ultra-Violet Optical Telescope (UVOT, \citealt{2005SSRv..120...95R}) onboard the \textit{Swift} satellite \citep{2004ApJ...611.1005G}.


The obtained data allow us to study the physical properties of the hosts of high-$A^{\rm GRB}_V$ GRBs in detail, and to investigate the bias against dust in GRB host samples. As an ultimate consequence they address the role of GRB host galaxies as tracers of galaxy formation and evolution. Furthermore, they link afterglow diagnostics, i.e., detailed information about a single sight line, to host-integrated properties, and in this combination directly probe the nature of dust and its properties in high-redshift, star-forming galaxies.

Throughout this work, we adopt the convention that the flux density of the afterglow $F_\nu(\nu, t)$ can be described as $F_\nu(\nu, t) \propto \nu^{-\beta}t^{-\alpha}$, and concordance ($\Omega_M=0.27$, $\Omega_{\Lambda}=0.73$, $H_0=71$~km/s/Mpc) $\Lambda$CDM cosmology. All errors are given at $1\sigma$ confidence unless indicated otherwise. All magnitudes and colors are given or converted into the AB system.

\section{Sample Selection}

\begin{figure}
\centering
\includegraphics[width=\columnwidth]{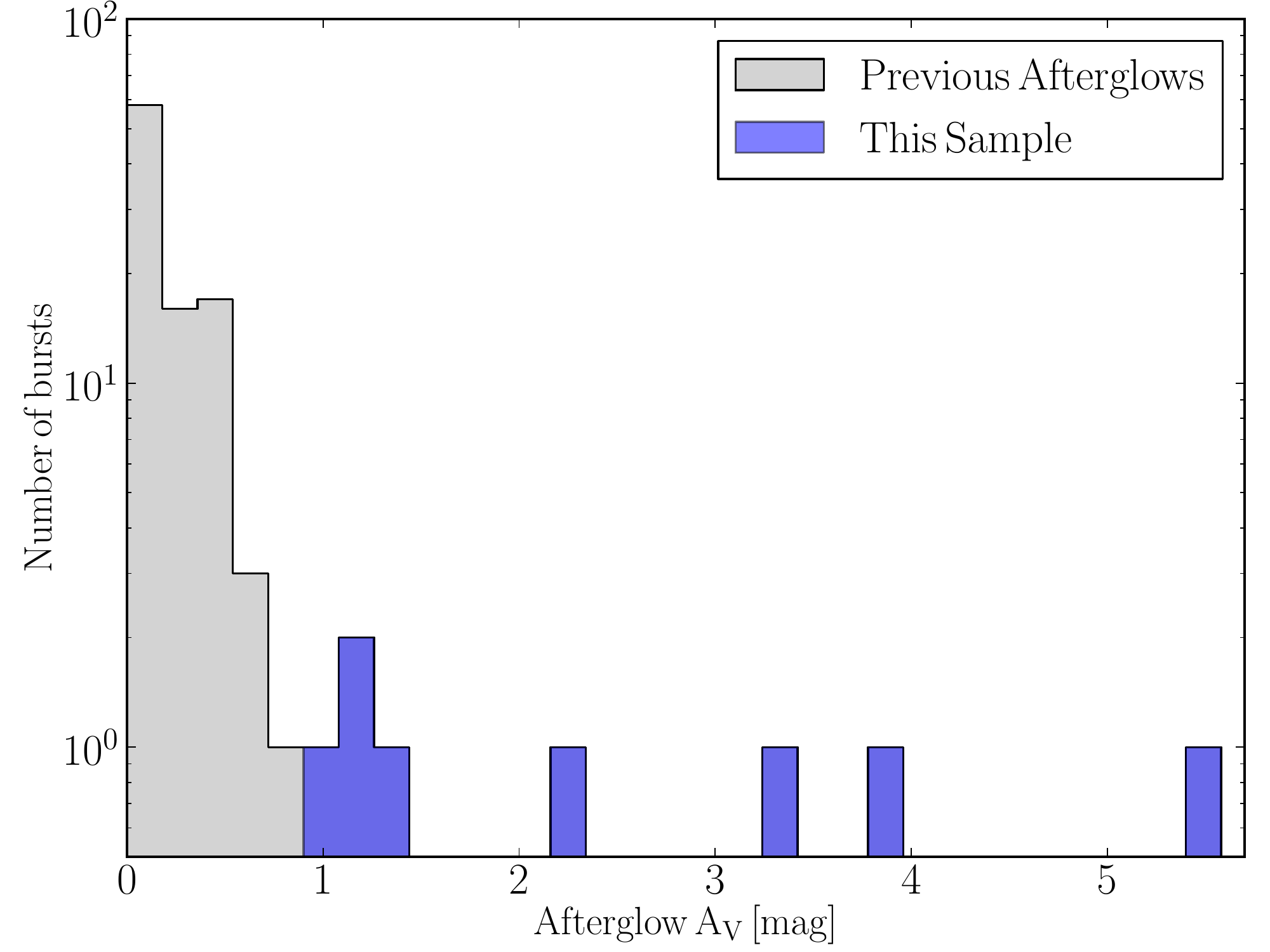}
\caption{Histogram of visual extinction values from previous afterglows and those selected for this study. We note that five of the afterglows in this work (GRBs~070802, 080605, 080607, 080805, 090926B) were already discussed in previous sample studies, but are only included in one of the histograms for clarity.}
\label{dusthist}
\end{figure}

The host galaxy sample presented in this work is based on a direct measurement of large visual extinction along the GRB line of sight ($A_V^{\rm GRB} \gtrsim 1\,\rm{mag}$) from multi-color (NIR to X-ray) afterglow observations. Specifically, eight GRB afterglows (GRBs~070306, 070802, 080605, 080607, 080805, 081109, 090926B and 100621A) fulfill the selection criterion and define our initial host sample. Our host sample is a direct result of afterglow observations. The selection itself is hence not limited by galaxy brightness, nor introduces a bias towards luminous galaxies. Afterglow measurements for the initial selection have been obtained from the literature or by analyzing photometric optical/NIR data from the GROND archive. In the latter case, they are detailed in Sect.~\ref{agobs}.

Our eight GRBs have a median redshift of $\langle z_{A_V} \rangle = 1.5$. This is significantly lower than the published mean of \textit{Swift} GRBs with measured redshifts ($\langle z_{Swift} \rangle = 1.9$, \citealt{2009ApJS..185..526F}\footnote{Including updates from http://www.mpe.mpg.de/\textasciitilde jcg/grbgen.html and http://www.raunvis.hi.is/\textasciitilde pja/GRBsample.html}), but larger than the one of the previous host sample ($\langle z_{\rm{SGL09}} \rangle = 0.96$ from SGL09), which includes a large number of pre-\textit{Swift} events. Within the selected eight afterglows, four of them show a clear 2175\AA~dust feature (GRBs 070802, 080605, 080607, 080805), which is with the exception of GRB~080603A \citep{2011arXiv1105.1591G, 2008arXiv0804.1959K} the full sample where a significant detection of this feature has been reported to date  \citep[e.g.,][]{2011arXiv1102.1469Z}. In fact, the 2175~\AA~feature could not have been detected in the rest of the sample. This arises from the combination of a large dust column, and not deep and rapid enough follow-up in the case of GRBs~070306 and 090926B, or no observational wavelength coverage  at $2175\,\AA \times(1+z)$ (GRBs 081109, 100621A).

While observationally challenging, the requirement of a dust measurement has several obvious advantages over a selection based on the optical-to-X-ray flux ratio ($\beta_{\rm oX}$, \citealp[see e.g.,][]{2004ApJ...617L..21J, 2005ApJ...624..868R, 2009ApJ...699.1087V}). Most importantly, the selection is the result of a measurement rather than an extrapolation, and provides a clean selection of dusty GRBs: Our afterglows are chosen accord to their visual extinction, instead of only their optical faintness. The latter events could of course also be at high-$z$ or have different emission mechanisms in the optical and X-ray regime.

We note, that our selection is still somewhat model-dependent, specifically that the afterglow emits synchrotron radiation in the optical/NIR and X-ray regime. Despite the lack of conclusive alternatives to the standard synchrotron-fireball model, there are still puzzling features in well-sampled multi-color light curves, such as chromatic\footnote{Chromatic light-curve breaks are breaks in the X-ray regime not associated with contemporanous breaks in the optical/NIR bands, and vice versa.} breaks, optical and/or X-ray flares or plateaus \citep[e.g.,][also illustrated in Fig.~\ref{100621lc}]{2006MNRAS.369.2059P, 2009MNRAS.397.1177E, 2009ApJ...697..758K, 2010arXiv1010.6212O}. The apparently decoupled optical and X-ray light curve for some bursts results in a strong dependence of $\beta_{\rm oX}$ on the time of the observation (which in fact is directly observed in some cases, see e.g., \citealt[][]{2011A&A...526A.113F}), and hence the dark burst definition depends also on observational constraints and not only physical properties. 

In addition, the required optical and/or NIR afterglow detection yields the GRB position to sub-arcsec accuracy, and hence negligible chance-coincidence probabilities for field galaxies, which is particularly important for small sample sizes as in this work. For all events a spectroscopic redshift is available from the literature, or could be obtained via host galaxy spectroscopy. This enables quantitative studies, such as the comparison between specific sight-lines against host-integrated properties, as well as an investigation of the relation between the dusty hosts to the host sample of SGL09.  

Due to the inherent difficulties to accurately localize high-$A_V^{\rm GRB}$ afterglows, most previous afterglow samples are biased towards small visual extinctions. This is illustrated in Fig.~\ref{dusthist}, which shows the visual extinction of GRBs selected for this work as compared to previously compiled $A_V^{\rm GRB}$ values. The latter were taken from \citet{2006ApJ...641..993K, 2010ApJ...720.1513K, 2010MNRAS.401.2773S, 2011A&A...526A..30G} and \citet{ 2011arXiv1102.1469Z}. 


\section{Observations}

\subsection{Afterglows}
\label{agobs}
Optical and near-infrared measurements of the afterglows of GRBs~070306, 070802, 080605, 080607, 080805 and 090926B or results thereof are taken from  \citet{2008ApJ...681..453J}, \citet{2008ApJ...685..376K}, \citet{2010arXiv1009.0004P}, \citet{2011A&A...526A..30G}, and \citet{2011arXiv1102.1469Z}, respectively. GROND observations of the afterglows of GRB 081109 and GRB 100621A are not presented elsewhere and are shortly described in the following.

\subsubsection{GRB~081109}

\textit{Swift} triggered on GRB~081109 \citep{2008GCN..8500....1I}, and X-ray and NIR measurements of the afterglow were rapidly reported by \citet{2008GCN..8506....1B} and \citet{2008GCN..8501....1D}. GROND observations in seven optical/NIR filters ($g^\prime r^\prime i^\prime z^\prime JHK_s$) simultaneously, started 17.1 hr after the GRB trigger \citep{2008GCN..8510....1C} and a preliminary analysis of the spectral energy distribution (SED) already revealed significant reddening of the optical/NIR afterglow \citep{2008GCN..8515....1C}. GROND continued to observe the transient at 2, 3, 6 and 378 days after the trigger, where the host brightness was derived from the last epoch. GROND afterglow and host measurements are given in Table~\ref{tab:agobs}, and Tables~\ref{tab:seds} and \ref{tab:sedsnir}, respectively. A possible host galaxy was also reported in the UVOT \emph{white} filter \citep{2008GCN..8523....1K}. High-energy prompt and afterglow data, early NIR imaging and a light curve and X-ray spectral analysis of this burst are presented by \citet{2009MNRAS.400.1829J}.

\begin{figure}
\centering
\includegraphics[height=\columnwidth, angle=-90]{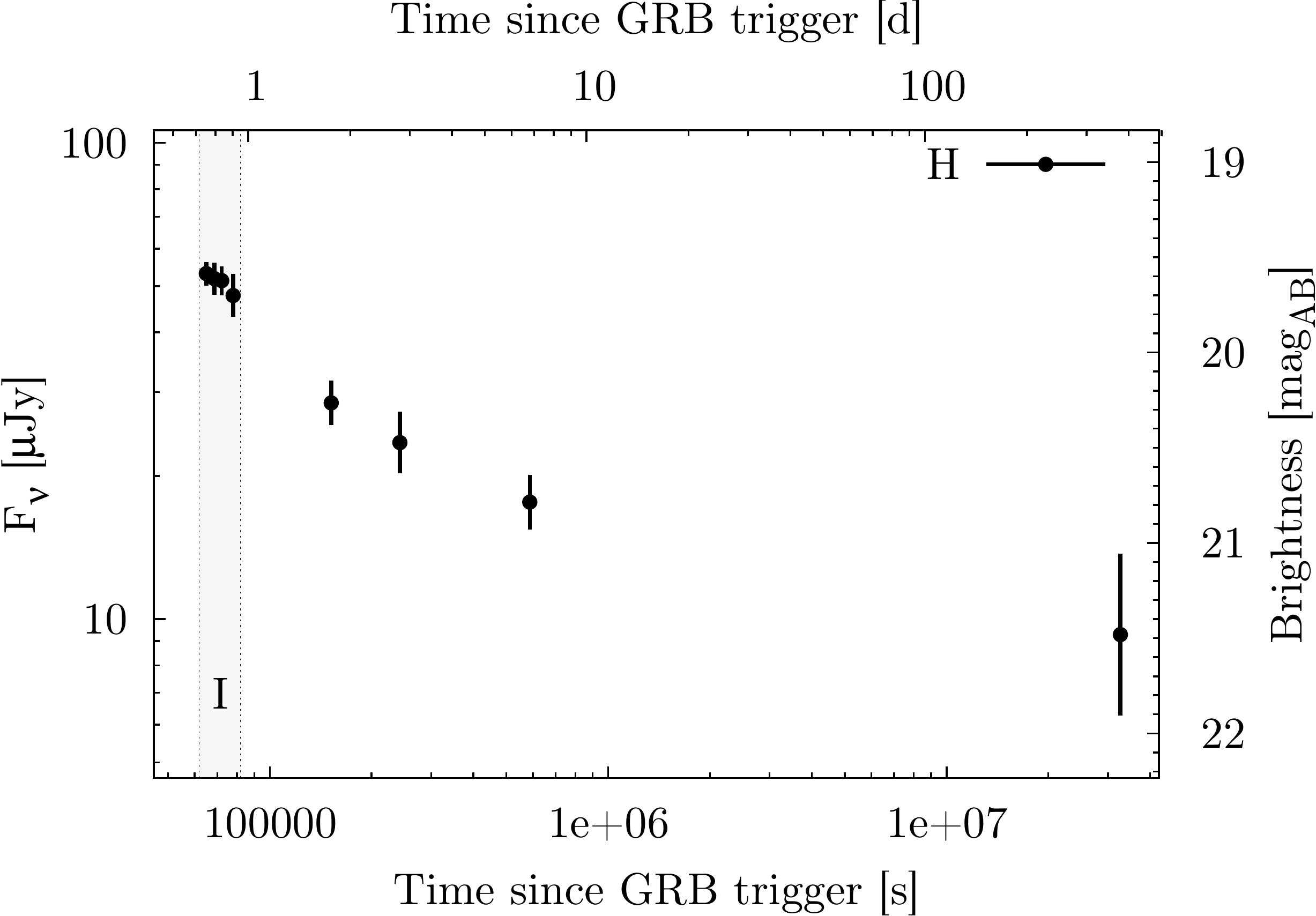}
\caption{GROND $H$-band light curve of the afterglow of GRB~081109. Highlighted with a grey shaded area and labeled with I is the time interval which has been used to extract a simultaneous, host subtracted afterglow SED from the GROND and \textit{Swift}/XRT data.}
\label{081109lc}
\end{figure}

\subsubsection{GRB~100621A}

\begin{figure}
\centering
\includegraphics[height=\columnwidth, angle=-90]{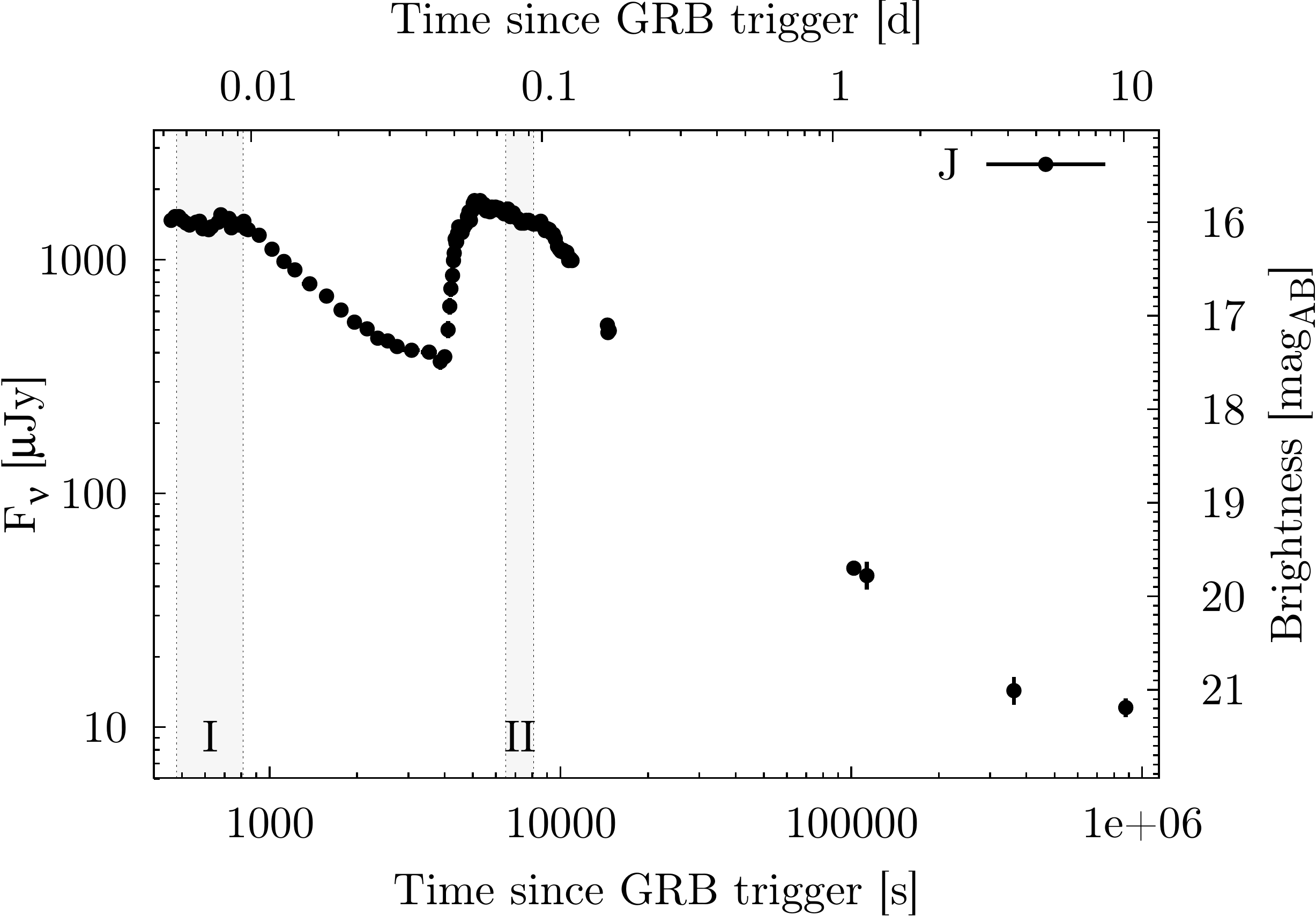}
\caption{GROND $J$-band light curve of the afterglow of GRB~100621A. Highlighted with grey shaded areas and labeled with I and II are the time intervals which have been used to extract a simultaneous, host subtracted afterglow SED from the GROND and \textit{Swift}/XRT data.}
\label{100621lc}
\end{figure}

GROND reacted immediately \citep{2010GCN.10874....1U} to the \textit{Swift} trigger of GRB~100621A \citep{2010GCN.10870....1U} taking the first images at 230~s after the burst. Simultaneous imaging in \gK continued for 3.05~h, and was resumed on night 2, 4 and 10 after the burst. Analysis of the \textit{Swift}/X-ray data and further detections of the NIR afterglow were reported by \citet{2010GCN.10877....1S} and \citet{2010GCN.10881....1N}, respectively. Early GROND data already revealed evidence for substantial reddening and host emission dominating the flux in the bluest filters, which was also seen by \textit{Swift}/UVOT \citep{2010GCN.10878....1O}. Afterglow and host measurements are again shown in Tables~\ref{tab:agobs},~\ref{tab:seds} and~\ref{tab:sedsnir}. The temporal evolution of the optical/NIR afterglow is complex with a very steep increase in brightness of around 1.5~mag in the $J$~band from 3.5 to 4.5~ks after the trigger. The light curve is hence very similar the one of GRB~081029 \citep{081029marco}, where an analogous behavior could be associated with the intrinsic properties of the GRB and not to changes in the intervening dust content. For completeness, the $J$-band light curve of GRB~100621A is shown in Fig.~\ref{100621lc}, but its detailed modeling and interpretation is beyond the scope of this paper, and will be discussed in a future work. 

\subsection{Hosts}

Once the sample based on afterglow host extinction was defined, late follow-up observations were initiated first with GROND, and in case of non-detections in individual filters, were continued with telescopes of successively increasing aperture size, specifically with EFOSC/SOFI at the NTT (4m class) and FORS2/HAWKI at the VLT (8m class). In one case without published redshift (GRB~081109), the photometric imaging was complemented by low-resolution spectroscopy with FORS2. Public VLT data for known hosts (GRBs 070306 and 070802) were obtained from the ESO archive. Ground-based data were complemented by \textit{Swift}/UVOT imaging for GRBs~081109 and 100621A. In the case of GRB~080607 all host measurements were taken from \citet{2010ApJ...723L.218C, 2011ApJ...727L..53C}. 

\section{Data reduction and SED fitting}

\subsection{Swift/XRT \& UVOT data}

X-ray data have been retrieved from the HEASARC archive\footnote{http://heasarc.gsfc.nasa.gov/cgi-bin/W3Browse/swift.pl}, and reduced with the \textit{xrtpipeline}. Spectra have been grouped to yield a minimum of 20 counts per bin, while the light curve was taken from the \textit{Swift}/XRT light curve repository \citep{2007A&A...469..379E, 2009MNRAS.397.1177E}. \textit{Swift}/UVOT photometry has been obtained following \citet{2008MNRAS.383..627P} and is provided in Table~\ref{tab:uvotseds}. 

\subsection{Ground-based optical/NIR photometry}

All optical (GROND, EFOSC, FORS1/2) and near-infrared (GROND, SOFI, HAWK-I, ISAAC) imaging was reduced in a standard manner using pyraf/IRAF \citep{1993ASPC...52..173T} similar to the procedure outlined in \citet{2008ApJ...685..376K}. For afterglow and host photometry, point-spread function (PSF) fitting and aperture photometry was used, respectively. The aperture diameter for individual hosts ranges between typically $1.0\arcsec$ to $2.5\arcsec$, corresponding to values between 1.5 to 4 times the PSF FWHM. It has been chosen sufficiently large to include the largest fraction of host flux, and given the typical extent of these galaxies ($\lesssim 1\arcsec$, i.e., $\lesssim 8.5\,\rm{kpc}$ at $z \sim 1.5$), the fraction of missed low-surface brightness emission is very likely not a major contribution to the presented measurements. 

PSF and aperture photometry were then flux calibrated against GROND observations of SDSS fields \citep{2009ApJS..182..543A} taken immediately before or after the GRB field for the optical \griz filters. $BVRI$ photometry was obtained by creating a set of $20-30$ secondary standards from the GROND photometry of field stars and the color terms from Lupton\footnote{http://www.sdss.org/dr7/algorithms/sdssUBVRITransform.html}. $U$-band photometry was tied to observations of Landolt standard stars \citep{1992AJ....104..340L} taken during the same night at different airmasses, which allowed a reliable correction of the atmospheric extinction to be applied. 

NIR photometry was measured against $20-60$ point sources from the 2MASS catalog \citep{2006AJ....131.1163S} in the $10\arcmin \times 10\arcmin$ field of view of GROND. The zeropoint for the NIR imagers with smaller field of views (in particular SOFI and ISAAC) was then derived using $3-10$ secondary standards common in the GROND and SOFI/ISAAC/HAWK-I frames. The HAWK-I $Y$-band imaging was calibrated against the $z^\prime$ and $J$ measurements from field stars using the color term
$$
Y = 0.05 + 0.463\times(z-J) + J
$$

where all magnitudes are in the AB system. This color term  has been obtained from synthetic photometry of stars with the templates of \citet{1998PASP..110..863P} and \citet{2000ApJ...542..464C}, and yields an rms residual scatter for individual stars of 0.07~mag. 

This procedure resulted in typical absolute accuracies of 2-5\% for the optical ($U$ to $z^\prime$) filters and 4-8\% in the NIR ($YJHK_s$), which have been added in quadrature to the error introduced by photon noise. All data used in the analysis have been corrected for the expected Galactic foreground extinction according to \citet{1998ApJ...500..525S} with $R_V=3.08$. For the selected fields, this correction is small, i.e., $E_{B-V} \lesssim 0.15\,\rm{mag}$ in all cases. 
The uncertainty of order of 10\% in the Galactic foreground correction is hence not going to affect the overall results of this work. Ground-based photometric measurements of the afterglows and hosts are shown in Tables~\ref{tab:agobs}, \ref{tab:seds} and ~\ref{tab:sedsnir}, respectively.

\begin{table*}
\caption{Afterglow photometric measurements \label{tab:agobs}}
\begin{tabular}{ccccccccc}
\hline
\noalign{\smallskip}
GRB & $\Delta$t [ks] & $g^\prime$ &  $r^\prime$ & $i^\prime$ & $z^\prime$ &$J$ & $H$ & $K_s$ \\  
\hline
081109 & 65 & $>$24.53 & $>$24.75 & 23.6(4) & 22.8(3) & 21.24(17) & 19.78(11) & 18.83(08) \\
100621A & 7.6 & 21.56(17) & 19.92(08) & 18.67(06) & 17.71(04) & 16.02(04) & 15.02(05) & 14.32(06) \\
\hline
\end{tabular}

\noindent{Notes: $\Delta$t is the mean time of the observation after the GRB trigger. All magnitudes are in the AB system and uncorrected for Galactic foreground reddening.  Values in brackets correspond to photometric errors in units of valid digits. Upper limits are given at 3$\sigma$ confidence.}
\end{table*}

\begin{table*}
\caption{\textit{Swift}/UVOT UV ($uvw2$ to $u$) photometric measurements of GRB hosts \label{tab:uvotseds}}
\begin{tabular}{ccccc}
\hline
\noalign{\smallskip}
Host & $uvw2$& $uvm2$& $uvw1$ & $u$ \\  
\hline
GRB~081109 & $>$24.1 &  23.6(3) & 23.4(3) & $>$22.9 \\
GRB~100621A & 22.31(04) & 22.23(06) & 22.20(07) & 21.95(06) \\
\hline
\hline
\end{tabular}

{Notes: All measurements in the AB system and uncorrected for the Galactic foreground reddening.  Values in brackets correspond to photometric errors in  in units of valid digits. Upper limits are given at 3$\sigma$ confidence.}
\end{table*}

\begin{table*}
\caption{Optical ($U$ to $z^\prime$) photometric measurements of GRB hosts \label{tab:seds}}
\begin{tabular}{ccccccccccc}
\hline
\noalign{\smallskip}
Host & $U$& $g^\prime$ & $V$ & $r^\prime$ & $R$ & $i^\prime$ & $I$ & $z^\prime$ \\  
\hline
GRB~070306 & --- & 22.90(09) & --- & 23.08(09) & 23.00(09) & 22.81(13)  & --- & 22.86(17) \\
GRB~070802 & --- & --- & --- & --- & 25.20(09) & 25.5(3)  & --- & --- \\
GRB~080605 & --- & 23.15(07) & --- & 22.82(07) & --- & 22.81(08) & --- & 22.76(11) \\
GRB~080805 & --- & --- & 25.7(2) & --- & 25.5(2) & 25.7(4) & --- & --- \\
GRB~081109 & 23.23(14) & 23.07(07) & 22.85(06) & 22.74(07) & --- & 22.01(08) & 21.96(09) & 21.99(09) \\
GRB~090926B & 23.71(13) &  23.31(07) & --- & 22.96(06) & --- & 22.92(12) & --- & 22.44(10) \\
GRB~100621A & 21.95(10) &  21.86(06) & --- & 21.48(06) & --- & 21.15(06) & --- & 21.46(06) \\
\hline
\hline
\end{tabular}

{Notes: All magnitudes in the AB system and uncorrected for the Galactic foreground reddening. Values in brackets correspond to photometric errors in units of valid digits. Upper limits are given at 3$\sigma$ confidence. Data for the host of GRB~080607 were taken from \citet{2010ApJ...723L.218C, 2011ApJ...727L..53C}, and are not shown in the table.}
\end{table*}

\begin{table*}
\caption{NIR ($Y$ to $K$) photometric measurements of GRB hosts \label{tab:sedsnir}}
\begin{tabular}{cccccccc}
\hline
\noalign{\smallskip}
Host & $Y$ & $J_{\rm{GROND}}$ & $J$ & $H_{\rm{GROND}}$ & $H$ & $K_{\rm{GROND}}$ & $K$ \\  
\hline
GRB~070306 & --- & 21.9(4) & 21.62(08) & 21.5(4) & 21.20(12) & $>$ 21.1 & 21.38(10) \\
GRB~070802 & --- & --- & 24.5(3) & --- & --- & --- & 23.4(3) \\
GRB~080605 & --- & 21.9(2) & --- & 22.3(3) & --- & $>$ 21.1 & ---  \\
GRB~080805 & --- & --- & 23.6(2) & --- & --- & ---  & 23.1(2) \\
GRB~081109 & 21.63(08) & 21.40(17) & 21.37(06) & 21.5(4) & --- & $>$ 20.6 & 21.05(08) \\
GRB~090926B & --- & 22.3(4) & 21.88(13) & $>$ 21.6 & 21.9(3) & $>$ 20.9 & 21.44(19) \\
GRB~100621A & 21.10(06) & 21.22(10) & 21.43(06) & 21.18(14) & --- & $>$ 21.1& 21.23(11)\\
\hline
\end{tabular}

{Notes: All measurements are given in the AB system and are uncorrected for the Galactic foreground reddening. Values in brackets correspond to photometric errors in units of valid digits. Upper limits are given at 3$\sigma$ confidence.}

\end{table*}

\subsection{Long-slit spectroscopy}

In addition to the photometric observations, the host of GRB~081109 was also observed spectroscopically with the VLT equipped with FORS2. In total $2\times1200$~s spectra were obtained with the grisms 300$V$ and 300$I$ and a long-slit width of $1.6\arcsec$. Acquisition images were taken in the $V$ and $I$ filter. The spectroscopic data were obtained at airmasses of $\sim1.1$ and seeing of $0.9\arcsec$, which results in a line-spread function of approximately $1.9$~nm at $570.0$~nm. The data were reduced using standard procedures in pyraf/IRAF, with the wavelength solution obtained against an HeHgCd arclamp exposures with 25 lines leaving residuals of around $0.07$~nm rms. Flux-calibration was performed against the spectro-photometric standard BPM16274\footnote{http://www.eso.org/sci/observing/tools/standards/spectra/bpm16274.html}. The wavelength- and flux-calibrated spectrum was corrected for Galactic foreground extinction 
and renormalized to the available photometry. 

\subsection{SED fitting}

\subsubsection{Afterglows}
\label{agsedfitt}
For the afterglow SED analysis, X-ray ($0.3-10$~keV) and optical/NIR data were fit together under the assumption that the underlying continuum emission is well represented by synchrotron emission \citep[e.g.,][]{1998ApJ...497L..17S, 2001ApJ...549L.209G, 2007MNRAS.377..273S, 2010arXiv1007.0383R}. 

Rest-frame soft X-ray photons are absorbed by metals, predominantly $\alpha$-chain elements, while UV over optical to NIR wavelengths are decreasingly affected by dust absorption. Good coverage above 1~keV combined with NIR observations allows for an accurate determination of the continuum, and hence good constraints on the dust abundance (represented by the absorption in the rest-frame, optical $V$-band, $A_V^{\rm GRB}$), the extinction law and total metal content (converted into a Hydrogen-equivalent column density $N_{H,X}$ assuming solar abundances from \citealt{1989GeCoA..53..197A}) along the line of sight. Specifically, single and broken power-law continua were used, where in the latter case the two power-law slopes $\beta_1$ and $\beta_2$ were fixed to yield $\beta_1+0.5=\beta_2$ as expected for the cooling break of synchrotron emission in the slow cooling regime \citep[e.g.,][]{1998ApJ...497L..17S, 2002ApJ...568..820G}. 

The NIR to X-ray SEDs were fit in X-spec \citep{1996ASPC..101...17A} using extinction laws for the Milky Way (MW), and Small (SMC) and Large (LMC) Magellanic Clouds from \citet{1992ApJ...395..130P}. For the GROND photometry, measurements from the time frame indicated in Fig.~\ref{081109lc} and ~\ref{100621lc} were used, where we chose interval II for GRB~100621A due to a better signal-to-noise ratio. Data taken at interval I yield consistent results for $A_V^{\rm GRB}$ and $N_{H,X}$. In the X-ray regime, where spectral changes in the late evolution of an afterglow are typically very moderate or in most cases even completely absent, the full XRT data set with a constant hardness ratio was used to create a time-averaged spectrum. The latter was then rescaled to the flux value at the time of the GROND observations derived from fitting the XRT light-curve with simple afterglow models. The early steep decay of the X-ray light curve and epochs of flaring activity were excluded from the combined spectrum as well as the light-curve fitting.

\subsubsection{Hosts}
\label{hostssed}
UV/optical/NIR photometry of the hosts of the selected GRBs were analyzed in a standard way using stellar population synthesis (SPS) techniques to convert luminosities into stellar masses $M_\ast$ \citep[e.g.,][]{2003ApJS..149..289B, 2010ApJ...709..644I} within LePhare\footnote{http://www.cfht.hawaii.edu/$\sim$arnouts/LEPHARE}. In detail, $3\times10^6$ galaxy templates based on models from \citet{2003MNRAS.344.1000B} with a universal IMF \citep{2003PASP..115..763C} and different ages, star formation histories, extinction laws, reddening values and metallicities were fit to the data. In addition, emission lines are taken into account by converting the de-reddened UV luminosity into a star formation rate, and hence line strengths of Ly-$\alpha$, $H_{\alpha}$, $H_{\beta}$, [OII] and [OIII] following \citet{1998ARA&A..36..189K} and \citet{2009ApJ...690.1236I}. In particular for vigorously star-forming galaxies such as GRB hosts, this effect is a significant contribution to even broad-band photometry \citep[e.g.,][]{2010arXiv1010.1783W} and reaches values of up to $\sim 0.2-0.3\,\rm{mag}$ \citep{2009ApJ...690.1236I}. For a direct comparison with results published in the literature \citep[e.g.,][SGL09]{2006A&A...459..745F, 2009ApJ...701.1765M, 2010ApJ...709..644I}, the attenuation law from \citet{2000ApJ...533..682C} derived for starburst galaxies is used, unless different reddening laws provide a better fit to the host data at 90\% confidence. 
We caution that access to the rest-frame NIR, which is the best tracer of the stellar mass of the galaxy, is somewhat limited for part of the sample. However, all hosts are detected in at least one filter redwards of the 4000~\AA~break, which allows a reasonable estimate of $M_\ast$ of a galaxy \citep[e.g.,][SGL09]{2004Natur.430..181G, 2009ApJ...690.1236I}.

\begin{figure}
\centering
\includegraphics[width=\columnwidth]{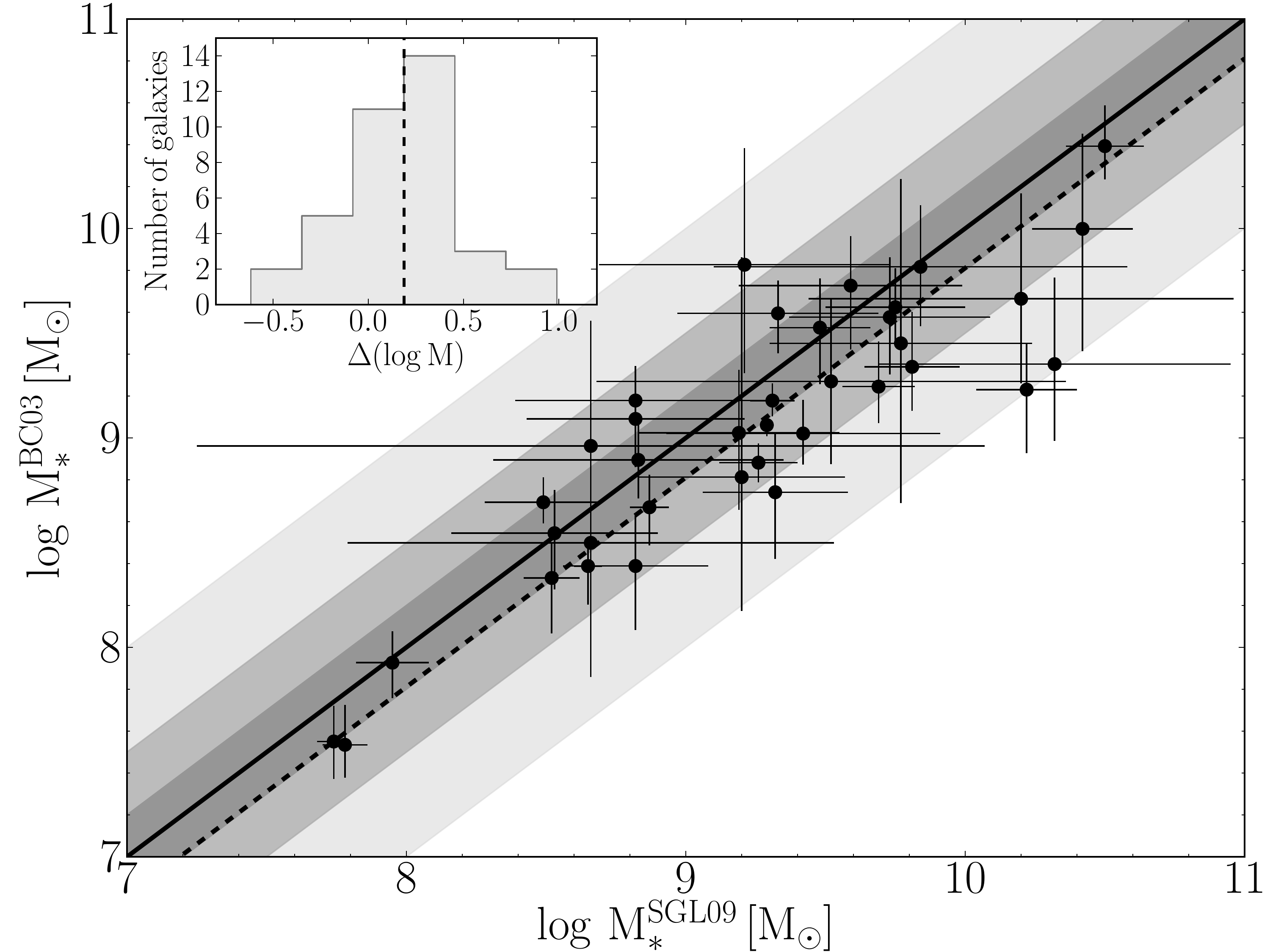}
\caption{Comparison of stellar masses from the host sample of SGL09. The plot shows host masses which were derived from the photometry compiled in SGL09 and following Sect.~\ref{hostssed} with models from \citet{2003MNRAS.344.1000B} against values directly taken from SGL09. Error bars are the maximum and minimum value of the stellar mass in the global $\chi^2$-distribution of $3\times10^6$ galaxy templates (see Sect.~\ref{hostssed}). The solid lines represents equality, and the dashed line the median offset. Increasingly grey shaded areas show dispersions of 0.2, 0.5 and 1.0 dex respectively. The inset shows the distribution of mass differences, which has a median of around 0.2~dex.}
\label{mass}
\end{figure}

Systematic uncertainties of up to an average of $0.2-0.3$~dex on $M_\ast$ are present due to the specific details of the stellar population models and the assumed attenuation/extinction law \citep{2007A&A...474..443P, 2007A&A...463..893K, 2009ApJ...702.1393K, 2010ApJ...709..644I}. Despite the small sample size of GRB hosts, there is evidence for an offset of around 0.2~dex between the presented method and the one of SGL09 as shown in Fig~\ref{mass}. In the following the recalculated stellar masses of the long GRB hosts with the photometric data compiled in SGL09 are used as a comparison sample.

The SPS fit returns not only the luminosity and mass of the galaxy, but also other physical properties of the host, such as the age of the dominant stellar population $\tau$, its color-excess $E_{(B-V)}$ and the star-formation rate (SFR) derived from the rest-frame UV flux. Reported physical host properties are the median of the probability distribution of the total grid over all galaxy templates at the fixed spectroscopic redshift. Errorbars represent the maximum and minimum value of the respective parameter in the global $\chi^2$-distribution of the multi-dimensional parameter grid. Typically, errors are asymmetric and dominated by the uncertainty of the color excess in the host galaxy, which results in logarithmic errors on all galaxy parameters. Absolute magnitudes and masses are compared against the redshift dependent galaxy luminosity functions \citep{2006ApJ...647..853W, 2007ApJ...656...42M}, and stellar mass function \citep{2009ApJ...701.1765M, 2010ApJ...709..644I}

\section{Results}

\subsection{Afterglow extinction and metals-to-dust ratios}
\label{agres}
\begin{table*}
\caption{Afterglow properties \label{tab:agres}}
\begin{tabular}{cccccccc}
\hline
\noalign{\smallskip}
Afterglow & Redshift$^{(*)}$ &  $T_{90}^{(**)}$ & $\beta_{\rm opt}^{(***)}$ & $\beta_{\rm X}$ & $A_V^{\rm GRB}$ & $N_{H, X}$ & $\chi^2$ / d.o.f. \\  
\hline
  &  & [s] &  &  & [mag] & $10^{22} \rm{cm}^{-2}$ &  \\  
\hline
GRB~070306  & 1.496  & 210$^{\rm{(a)}}$ & 1.00 & $1.00\pm0.07$ & $5.5^{+1.2}_{-1.0}$ & $2.5_{-0.2}^{+0.3}$ & 123 / 108 \\
GRB~070802$^{\rm{(b,c)}}$  & 2.452  & 16.4$^{\rm{(d)}}$ & 0.60 & $1.10_{-0.12}^{+0.14}$ & $1.23^{+0.18}_{-0.16}$ / $1.19\pm0.15$ & $2.0_{-0.8}^{+0.7}$ / $<2.9$ & 15 / 14 \\
GRB~080605$^{\rm{(b,c)}}$  & 1.640  & 20$^{\rm{(e)}}$ & 0.67 & $0.67\pm0.01$ & $0.47\pm0.03$ / $1.20_{-0.10}^{+0.09}$ & $1.01_{-0.08}^{+0.09}$ / $0.71\pm0.08$ & 428 / 327 \\
GRB~080607$^{\rm{(c,f)}}$  & 3.036  & 79$^{\rm{(g)}}$ & $0.96$ & $0.96_{-0.06}^{+0.05}$ & $3.3\pm0.3$ / $2.33_{-0.43}^{+0.46}$ & $2.7_{-0.7}^{+0.8}$ / $3.8_{-0.2}^{+0.2}$ & 70 / 39\\
GRB~080805$^{\rm{(b,c)}}$  & 1.505 & 78$^{\rm{(h)}}$ & $0.47$ & $0.97\pm0.05$ & $1.01^{+0.19}_{-0.14}$ / $1.53\pm0.13$ & $1.0_{-0.4}^{+0.6}$ / $1.2_{-0.5}^{+0.4}$ & 23 / 18 \\
GRB~081109  & 0.979  & 190$^{\rm{(i)}}$ &  1.10 & $1.12_{-0.02}^{+0.02}$ & $3.4^{+0.4}_{-0.3}$ & $1.10_{-0.12}^{+0.13}$ & 48 / 63 \\
GRB~090926B$^{\rm{(b)}}$ & 1.24  & 110$^{\rm{(j)}}$ & 0.73 & $0.73_{-0.07}^{+0.09}$ & $1.4_{-0.6}^{+1.1}$ & $2.2_{-0.4}^{+0.5}$ & 33 / 31 \\
GRB~100621A & 0.542 & 63.6$^{\rm{(k)}}$ &  0.79 & $1.29_{-0.10}^{+0.11}$ & $3.8^{+0.2}_{-0.2}$ & $1.62_{-0.15}^{+0.15}$ & 138 / 124 \\

\hline
\hline
\end{tabular}

\noindent{$^{(*)}$ Redshifts from \citet{2008ApJ...681..453J}, \citet{2009ApJ...697.1725E}, \citet{2009ApJS..185..526F}, \citet{2009GCN..9947....1F} and \citet{2010GCN.10876....1M}. \\
$^{(**)}$ $T_{90}$ is the duration in which the GRB emits from 5\% to 95\% of its $\gamma$-rays, and is used to discriminate between short and long bursts. Typically, long GRBs have $T_{90} > 2~s$ \citep{1993ApJ...413L.101K}. All GRBs in this work are hence unambigously long events.\\
$^{(***)}\beta_{\rm opt}$ is tied to $\beta_X$ in the fitting, and hence has the same error.

Notes: Afterglow measurements were taken from the reference denoted with the superscript in the first line. When two references are given, we quote both values for $A_V^{\rm GRB}$ and $N_{H, X}$, but values for $\beta$ and $\chi^2$ only from the first one for the sake of clarity.

References: (a) \citet{2007GCN..6173....1B}, (b) \citet{2011A&A...526A..30G}, (c) \citet{2011arXiv1102.1469Z}, (d) \citet{2007GCN..6699....1C}, (e) \citet{2008GCN..7841....1C}, (f) \citet{2010arXiv1009.0004P}, (g) \citet{2008GCN..7852....1S}, (h) \citet{2008GCN..8068....1P}, (i) \citet{2008GCN..8507....1M}, (j) \citet{2009GCN..9939....1B}, (k) \citet{2010GCN.10875....1U}
}
\end{table*}

Out of the total of eight afterglows in the sample, the extinction properties for five of them (GRBs 070802, 080605, 080607, 080805, 090926B) are extensively discussed in previous works \citep{2008ApJ...685..376K, 2009ApJ...697.1725E, 2010arXiv1009.0004P, 2011A&A...526A..30G, 2011arXiv1102.1469Z}. In these cases, also the published afterglow analysis is comparable to the approach of this work, and it is therefore not repeated here, with results from the literature being summarized in Table~\ref{tab:agres}. For the remaining three events (GRBs~070306, 081109, 100621A), we present either new data and their modeling (GRBs 081109, 100621A), or a new analysis (GRB~070306) in the following.

\begin{figure}
\centering
\includegraphics[width=\columnwidth]{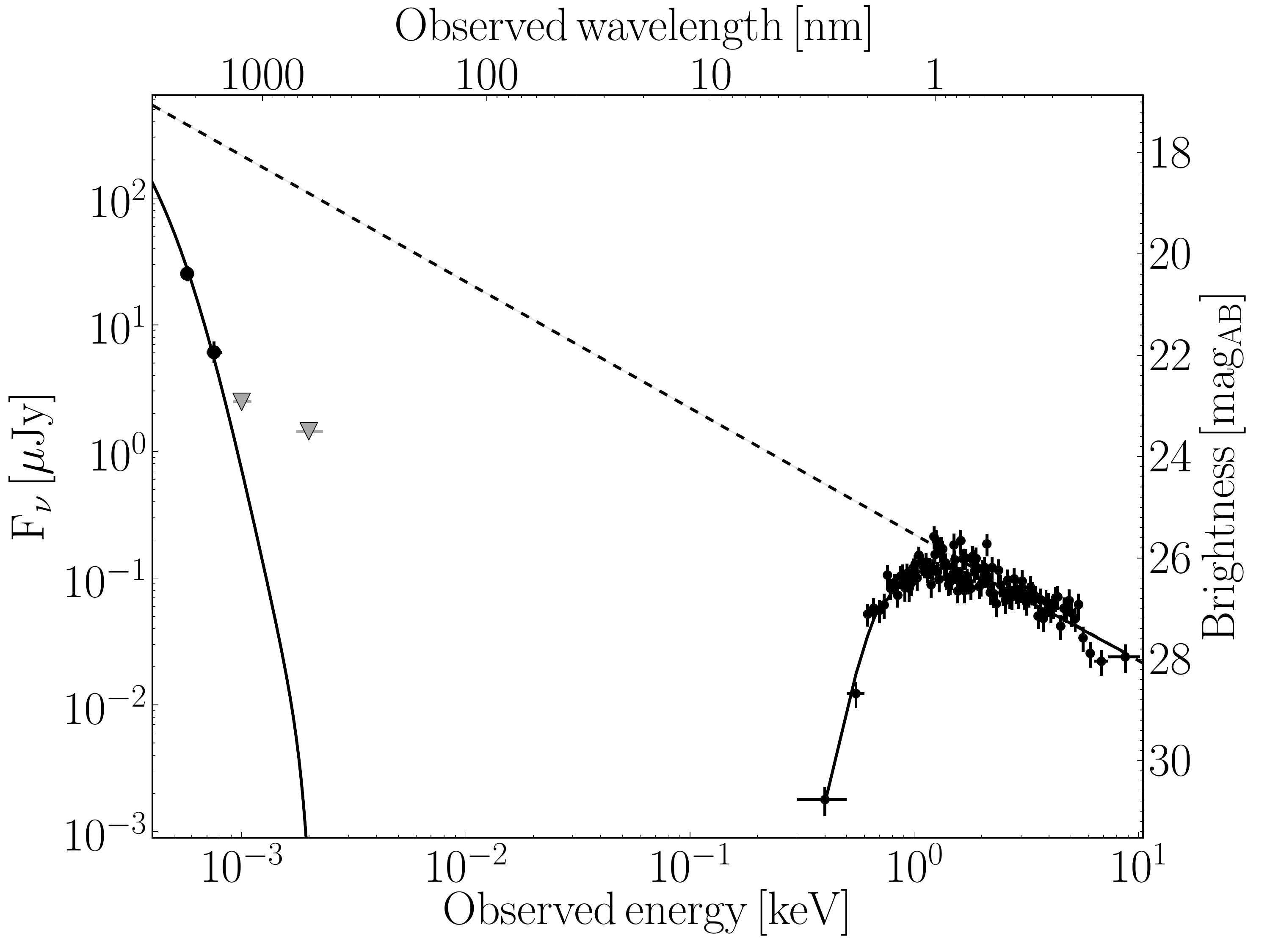}
\caption{NIR to X-ray SED of the afterglow of GRB~070306 obtained approximately 120~ks after the trigger in the observer's frame. The dashed line shows the unabsorbed synchrotron continuum emission while the best-fit model (including dust and metal absorption) is shown by solid lines. Upper limits are shown by downward triangles.}
\label{070306sed}
\end{figure}

The afterglow of \textbf{GRB~070306} was discussed in \citet{2008ApJ...681..453J}. As their analysis is significantly different from our approach, we refit the available afterglow data following Sect. \ref{agsedfitt}. The broad-band SED in Fig.~\ref{070306sed} is reasonably  well fit ($\chi^2 = 123/108$ d.o.f.) with a single power-law continuum with spectral index $\beta = 1.00\pm 0.07$, an $A_V^{\rm GRB} = 5.5_{-1.0}^{+1.2}\,\rm{mag}$ and $N_{H,X} = 2.5_{-0.2}^{+0.3}\times 10^{22}$cm$^{-2}$ at 90\% confidence, which implies a metals-to-dust ratio of $N_{H,X}/A_V^{\rm GRB} = 4.4\times10^{21}$cm$^{-2}$/mag. Given the redshift of $z=1.496$, and the sparse wavelength coverage in the NIR (probing the rest-frame optical red-ward of 400~nm, where there is little distinction between local extinction laws), all local dust models provide equally good fits to the data of course, and within errors compatible values of $\beta$, $A_V^{\rm GRB}$, and $N_{H,X}$. No strong statements can be made either with respect to a possible presence of a break between NIR and X-ray data. We adopt the model with the least number of free parameters (single power-law continuum, which also yields the lowest $\chi^2$), but note that in the case of a break between the two wavelength regimes (as seen in most early GRB afterglows, \citealt{2011A&A...526A..30G}) the fit is of comparable quality ($\Delta\chi^2=3$), and yields an best-fit $A_V^{\rm GRB}$ which would be significantly lower ($A_V^{\rm GRB} = 4.3_{-1.0}^{+1.1}\,\rm{mag}$), but within errors consistent with the single power-law values.

\begin{figure}
\centering
\includegraphics[width=\columnwidth]{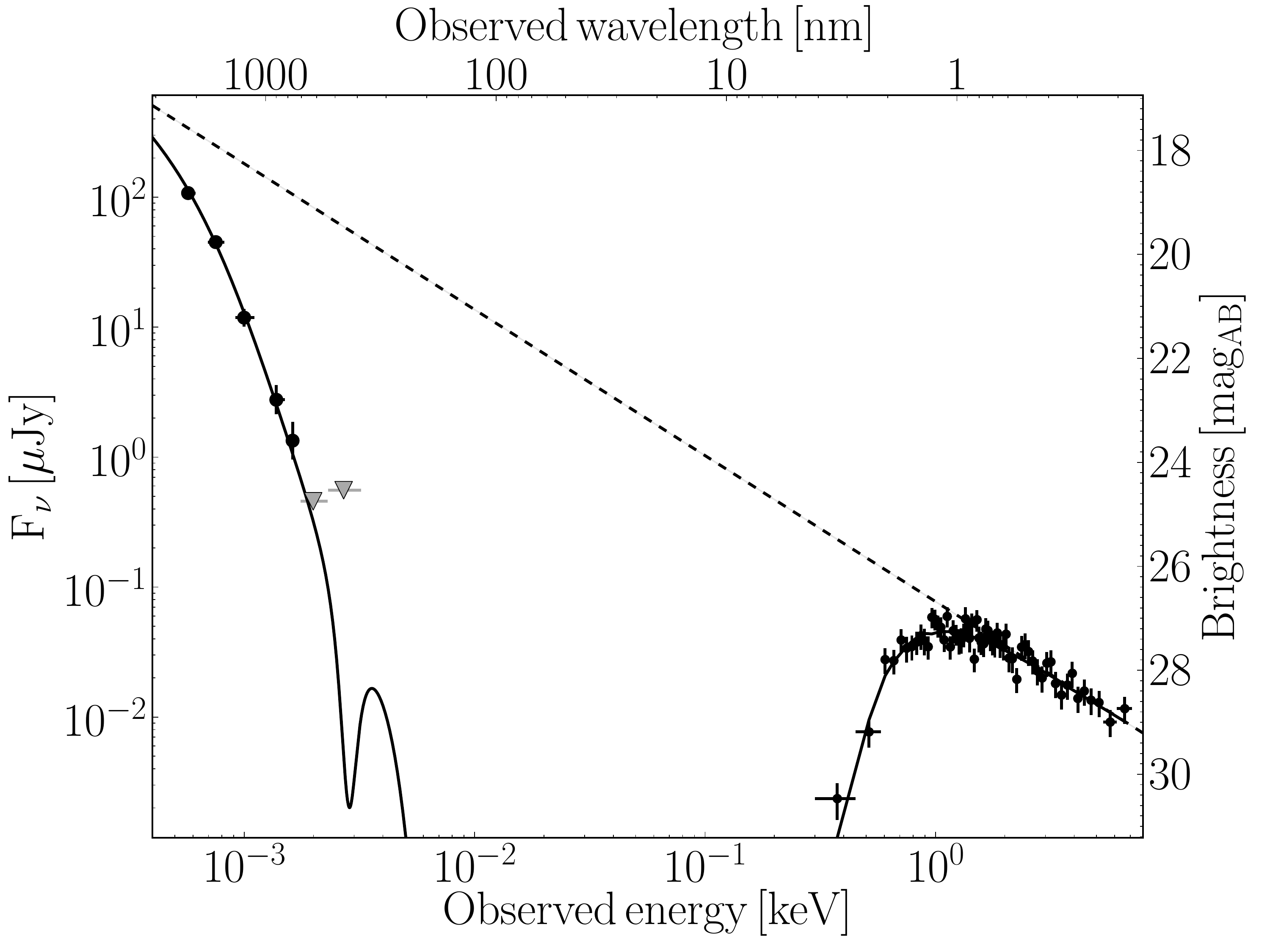}
\caption{Same as Fig.~\ref{070306sed} for the afterglow of GRB~081109 obtained approximately 60~ks after the trigger.}
\label{081109sed}
\end{figure}

The afterglow SED for \textbf{GRB~081109} at $z=0.979$ has been constructed from GROND and \textit{Swift} data and is shown in Fig.~\ref{081109sed}. After subtraction of the late host epoch, no residual flux is detected in the two bluest GROND filters $g^\prime$ and $r^\prime$. The afterglow is detected in all five redder bands, which implies an extremely red color of $(R-K)_{\rm AB}\gtrsim6\,\rm{mag}$ and $\beta_{\rm oX} < 0.44$. The combined GROND and XRT data are well fit with a single power-law continuum, indicating that both the optical/NIR and the X-ray regimes probe the same part of the synchrotron spectrum. The obvious curvature in the GROND data is accuratly described with either of the local dust models, with best-fit parameters of $A_V^{\rm GRB} = 3.4_{-0.3}^{+0.4}\,\rm{mag}$, $\beta = 1.12\pm0.02$, and $N_{H.X} = (1.1\pm0.1)\times 10^{22}$cm$^{-2}$ at 90\% confidence and a $\chi^2$ of 48.1 for 63~d.o.f. in a MW-like extinction law. SMC and LMC models yield within errors comparable parameters and provide equally good fits to the data ($\chi^2$s of 48.2 and 48.8, respectively). 

\begin{figure}
\centering
\includegraphics[width=\columnwidth]{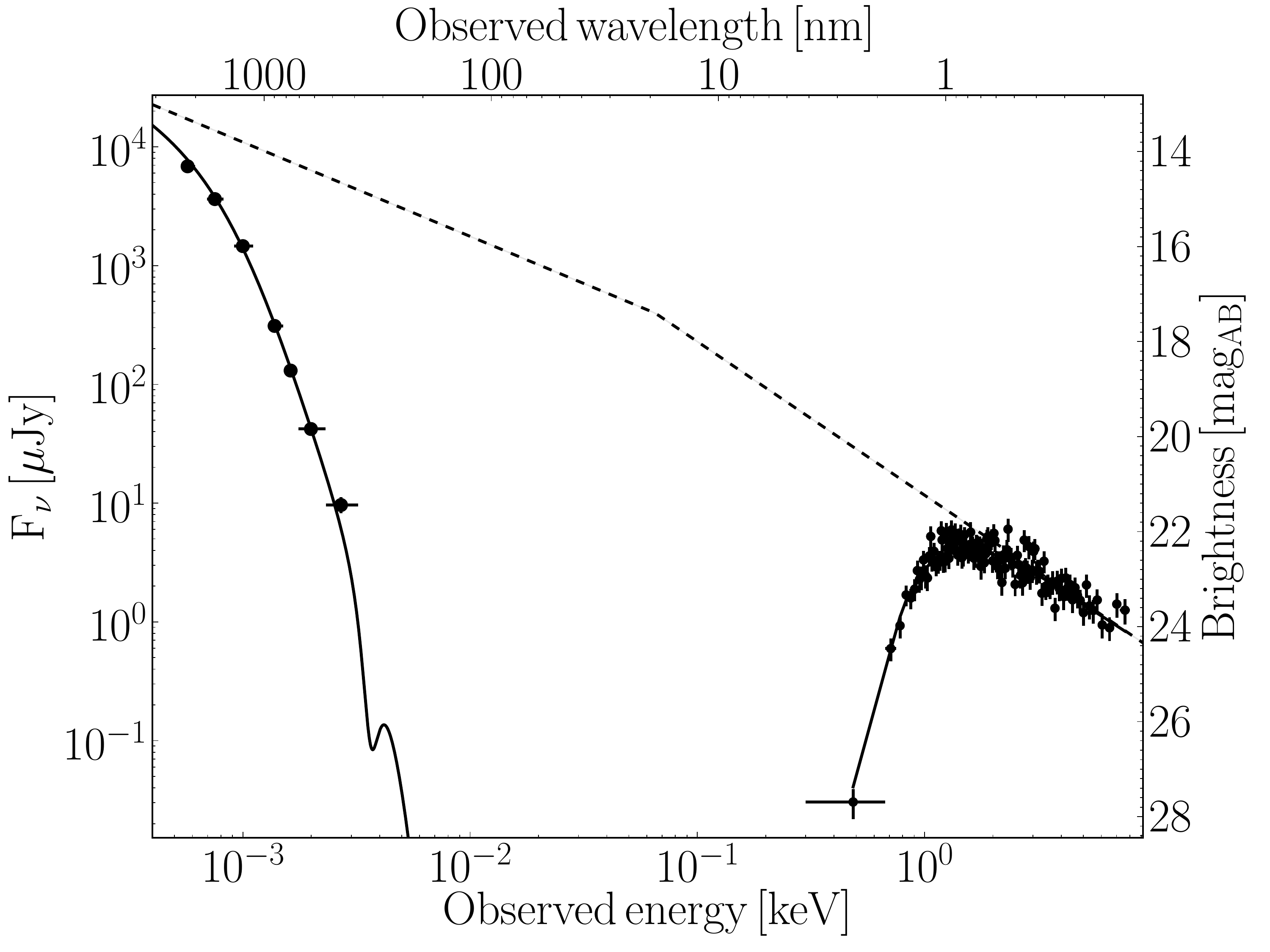}
\caption{Same as Fig.~\ref{070306sed} for the afterglow of GRB~100621A obtained approximately 7.6~ks after the trigger.}
\label{100621Ased}
\end{figure}

The SED of the afterglow of \textbf{GRB~100621A} at $z=0.542$ \citep{2010GCN.10876....1M} is shown in Fig.~\ref{100621Ased}. Similarly to the SED of GRB~081109 there is strong curvature and obvious reddening in the optical/NIR part of the SED. The inferred ultra-red color of $(R-K)_{\rm AB} \sim$~5.8~mag, and the $\beta_{\rm oX}$ value of 0.39 provide evidence for strong dust extinction. The best fit is obtained with a broken power-law with spectral indices $\beta_1 = \beta_2 - 0.5 = 0.79^{+0.11}_{-0.10}$, as well as $A_V^{\rm GRB} = 3.8\pm0.2\,\rm{mag}$ for an LMC-like extinction law, and $N_{H.X} = (1.62\pm0.15)\times 10^{22}$cm$^{-2}$ at 90\% confidence ($\chi^2$ of 138 for 124 d.o.f.). Given the rest-frame coverage of $\sim300-1500$~nm all local dust models return comparable values with $A_V^{\rm GRB}$ values of $3.8\pm0.2\,\rm{mag}$ for an SMC- and $4.0\pm0.2\,\rm{mag}$ for a MW-type extinction law. 
All data bluewards and including the $r^\prime$ filter are consistent with this extinction laws, while the $g^\prime$-band photometry is somewhat ($2-3\sigma$) brighter than the best fit predicts. This could indicate a discrepancy between the details of the specific dust extinction law and local models similar as observed e.g., in GRBs~070802 or 080607 \citep{2009ApJ...697.1725E, 2010arXiv1009.0004P}.

The visual extinctions for GRBs~070306, 081109 and 100621A are among the largest ever measured directly along GRB sight-lines, and further imply metals-to-dust ratio of $N_{H,X}/A_V^{\rm GRB} \sim (3-5)\times10^{21}$cm$^{-2}$/mag. Compared to previous afterglow measurements, these $N_{H,X}/A_V^{\rm GRB}$ values are relatively low, and within a factor of 2-3 similar to $N_{H}/A_V$ as observed in the SMC, LMC and MW (see also Sect. \ref{avnhcon}). The results of afterglow SED fitting, as well as values taken from the literature, are summarized in Table~\ref{tab:agres}.

\subsection{Host properties}
\label{hostres}
\begin{table*}
\caption{Host galaxy properties \label{tab:hostprop}}
\begin{tabular}{ccccccccc}
\hline
\noalign{\smallskip}
Host & $z_{\rm spec}$ & $z_{\rm phot}$  &  R.A.$^{(a)}$ & Decl.$^{(a)}$ & Offset$^{(b)}$ & Probability$^{(c)}$ & Instruments$^{(d)}$ & $Filter$\\  
\hline
  &  &  & J2000 & J2000 &  arcsec &  &  & \\  
\hline
GRB~070306 & 1.496 & $1.64_{-0.13}^{+0.35}$ & 09:52:23.31 & +10:28:55.4 & $\lesssim 0\farcs{2}$ & 0.002 & FGI & $g^\prime r^\prime Ri^\prime z^\prime J_GJHK^{(f)}$ \\
GRB~070802 & 2.452 & $2.3_{-0.6}^{+1.5}$ & 02:27:35.72 & -55:31:39.0 & $0\farcs{15}\pm0\farcs{08}$ & 0.011 & EFHI & $RIJK$ \\
GRB~080605 & 1.640 & $1.7_{-0.2}^{+0.7}$ & 17:28:30.05 & +04:00:56.0 & $0\farcs{23}\pm0\farcs{11}$ & 0.002 & G & $g^\prime  r^\prime i^\prime z^\prime JH$ \\
GRB~080805 & 1.505 & $1.7_{-0.7}^{+0.6}$ & 20:56:53.43 & -62:26:39.3 & $\lesssim 0\farcs{3}$ & 0.030 & EH & $VRIJK$ \\
GRB~081109 & 0.979 & $0.84^{+0.16}_{-0.07}$ & 22:03:09.63 & -54:42:39.9 & $\lesssim 0\farcs{2}$ & 0.002 & EFGHU & $Ug^\prime Vr^\prime i^\prime Iz^\prime YJ_GJHK$ \\
GRB~090926B & 1.24 & $1.43^{+0.62}_{-0.26}$ & 03:05:13.91 & -39:00:22.6 & $\lesssim 0\farcs{6}$ & 0.018 & EFGS & $Ug^\prime r^\prime i^\prime z^\prime JHK$ \\
GRB~100621A & 0.542 & $0.50^{+0.14}_{-0.02}$ & 21:01:13.08 & -51:06:22.2 &  $0\farcs{12}\pm0\farcs{08}$ & 0.0006 & EGHU & $w2m2w1uUg^\prime r^\prime i^\prime z^\prime YJHK$ \\
\hline
\hline
\end{tabular}

\noindent{Notes: Redshifts from \citet{2008ApJ...681..453J}, \citet{2009ApJ...697.1725E}, \citet{2009ApJS..185..526F}, \citet{2009GCN..9947....1F} and \citet{2010GCN.10876....1M}. The host of GRB~080607 is not shown in this table. All measurements were directly taken from \citet{2010ApJ...723L.218C, 2011ApJ...727L..53C}

$^{(a)}$ Host position derived after tying the astrometric solution to the USNO-B1 catalog \citep{2003AJ....125..984M}. Typical absolute uncertainties are $\approx 0\farcs{3}$.

$^{(b)}$ Relative offset calculated by registering the host images against astrometric templates derived from afterglow images with a typical precision of 40 mas rms 

$^{(c)}$ Estimated chance coincidence probability following \citet{2002AJ....123.1111B} and \citet{2009AJ....138.1690P}

$^{(d)}$ G is GROND at the 2.2~m MPG/ESO telescope, E/S are EFOSC/SOFI at the NTT, and F/H/I are HAWK-I, FORS1/2 and ISAAC at the VLT, and U UVOT onboard \textit{Swift} respectively. 

$^{(f)}$ Further complemented by the $u$ and $I$ band magnitudes in \citet{2007GCN..6170....1C} and \citet{2008ApJ...681..453J}. 

 }

\end{table*}

\begin{table*}
\caption{SPS Host galaxy fitting results\label{tab:hostres}}
\begin{tabular}{cccccccccc}
\hline
\noalign{\smallskip}
Host & Redshift &  M$_B$ & Age & $E'_{B-V}$ & $\log (M_\ast)$ & $\log$ SFR & $\log$ SSFR & L & $\chi^2$/N$_{\rm filt}$ \\  
\hline
  &  &  mag$_{\rm AB}$ & Gyr & mag & \Msun & \Msun /yr & yr$^{-1}$ & $L^{*}_B$ & \\  
\hline
GRB~070306  & 1.496 & $-22.4\pm0.1$ & $1.6^{+1.6}_{-1.0}$ & $<0.16$ (2$\sigma$) & $10.39^{+0.19}_{-0.15}$ & $1.1^{+0.3}_{-0.2}$ & $-9.3^{+0.4}_{-0.3}$ & $1.7\pm0.2$ & 11.8/12 \\
GRB~070802  & 2.452 & $-21.4\pm0.2$ & $0.38^{+0.66}_{-0.31}$ & $<0.42$ (2$\sigma$) & $9.7^{+0.2}_{-0.3}$ & $1.0^{+0.6}_{-0.5}$ & $-8.7^{+1.0}_{-0.6}$ & $0.6\pm0.2$ & 2.3/4 \\
GRB~080605  & 1.640 & $-22.6\pm0.2$ & $0.06^{+0.18}_{-0.03}$ & $<0.22$ (2$\sigma$) & $9.6^{+0.3}_{-0.2}$ & $1.6^{+0.3}_{-0.3}$ & $-8.0^{+0.5}_{-0.6}$ & $2.1\pm0.4$ & 3.3/6 \\
GRB~080607  & 3.036 & $-21.1\pm0.1$ & $0.16^{+0.53}_{-0.12}$ & $0.35^{+0.25}_{-0.09}$ & $9.9^{+0.4}_{-0.6}$ & $1.6^{+0.4}_{-0.5}$ & $-8.2^{+0.6}_{-0.7}$ & $0.3\pm0.1$ & 3.9/5 \\
GRB~080805  & 1.505 & $-20.4\pm0.2$ & $0.51^{+0.74}_{-0.40}$ & $<0.65$ (2$\sigma$) & $9.7^{+0.2}_{-0.2}$ & $0.8^{+0.7}_{-0.8}$ & $-8.9^{+0.9}_{-0.9}$ & $0.3\pm0.1$ & 0.3/5 \\
GRB~081109  & 0.979 & $-21.27\pm0.09$ & $0.24^{+0.22}_{-0.11}$ & $0.24_{-0.04}^{+0.06}$ & $9.82^{+0.09}_{-0.09}$ & $1.5^{+0.2}_{-0.2}$ & $-8.4^{+0.3}_{-0.3}$ & $0.9\pm0.1$ & 11.2/14 \\
GRB~090926B & 1.24  & $-21.5\pm0.1$ & $0.14^{+0.60}_{-0.09}$ & $0.35_{-0.05}^{+0.08}$ & $10.1^{+0.4}_{-0.5}$ & $1.9^{+0.4}_{-0.5}$ & $-8.1^{+0.6}_{-0.9}$ & $0.9\pm0.1$ & 7.3/9 \\
GRB~100621A & 0.542 & $-20.68\pm0.08$ & $0.05^{+0.07}_{-0.03}$ & $0.14_{-0.04}^{+0.03}$ & $8.98^{+0.14}_{-0.10}$ & $1.13^{+0.15}_{-0.20}$ & $-7.9^{+0.2}_{-0.3}$ & $0.6\pm0.1$ & 16.5/14 \\

\hline
\hline
\end{tabular}
\end{table*}

\subsubsection{The host of GRB 070306}

\begin{figure*}
\centering
\includegraphics[width=.9\columnwidth]{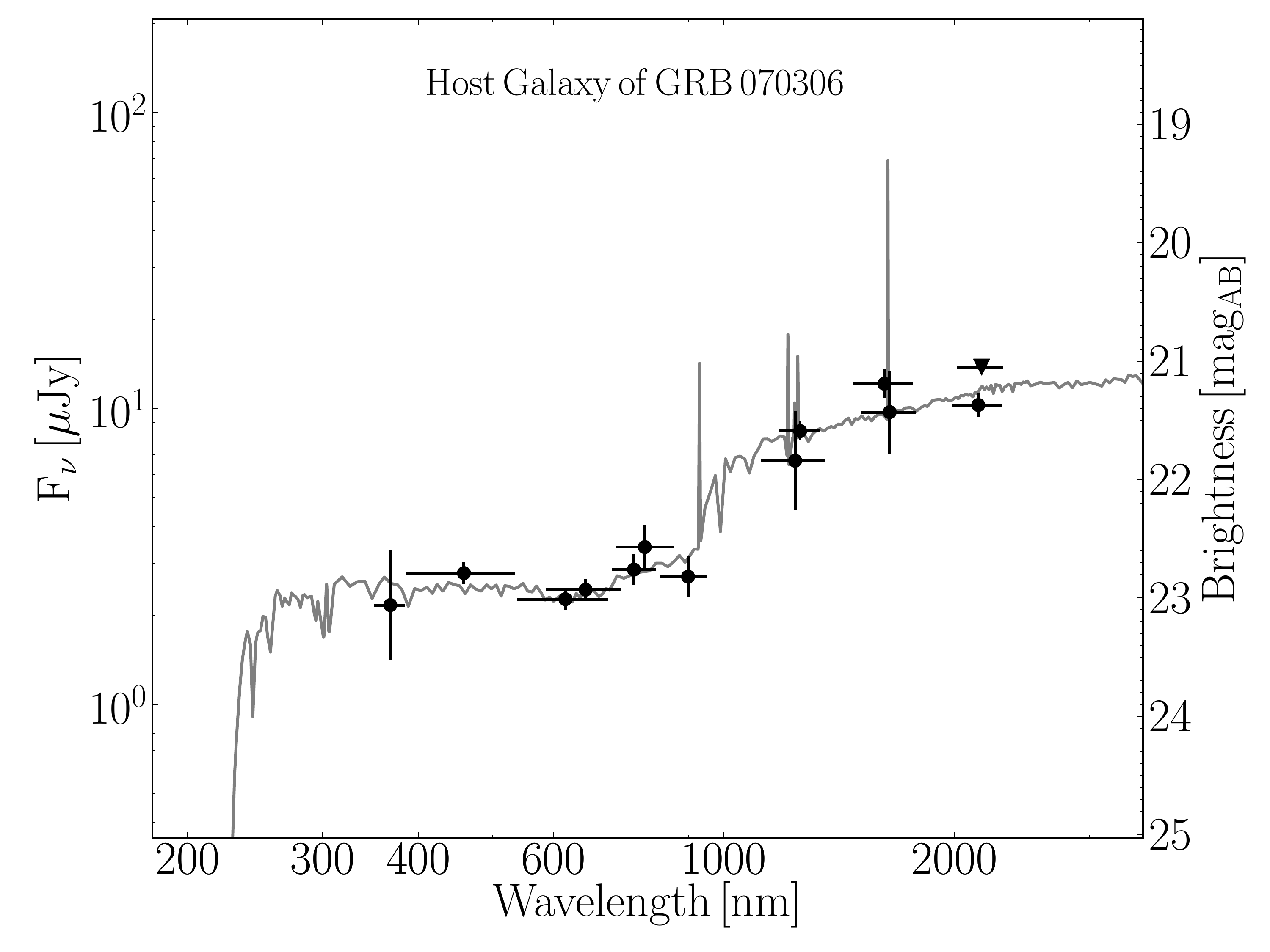}
\includegraphics[width=.9\columnwidth]{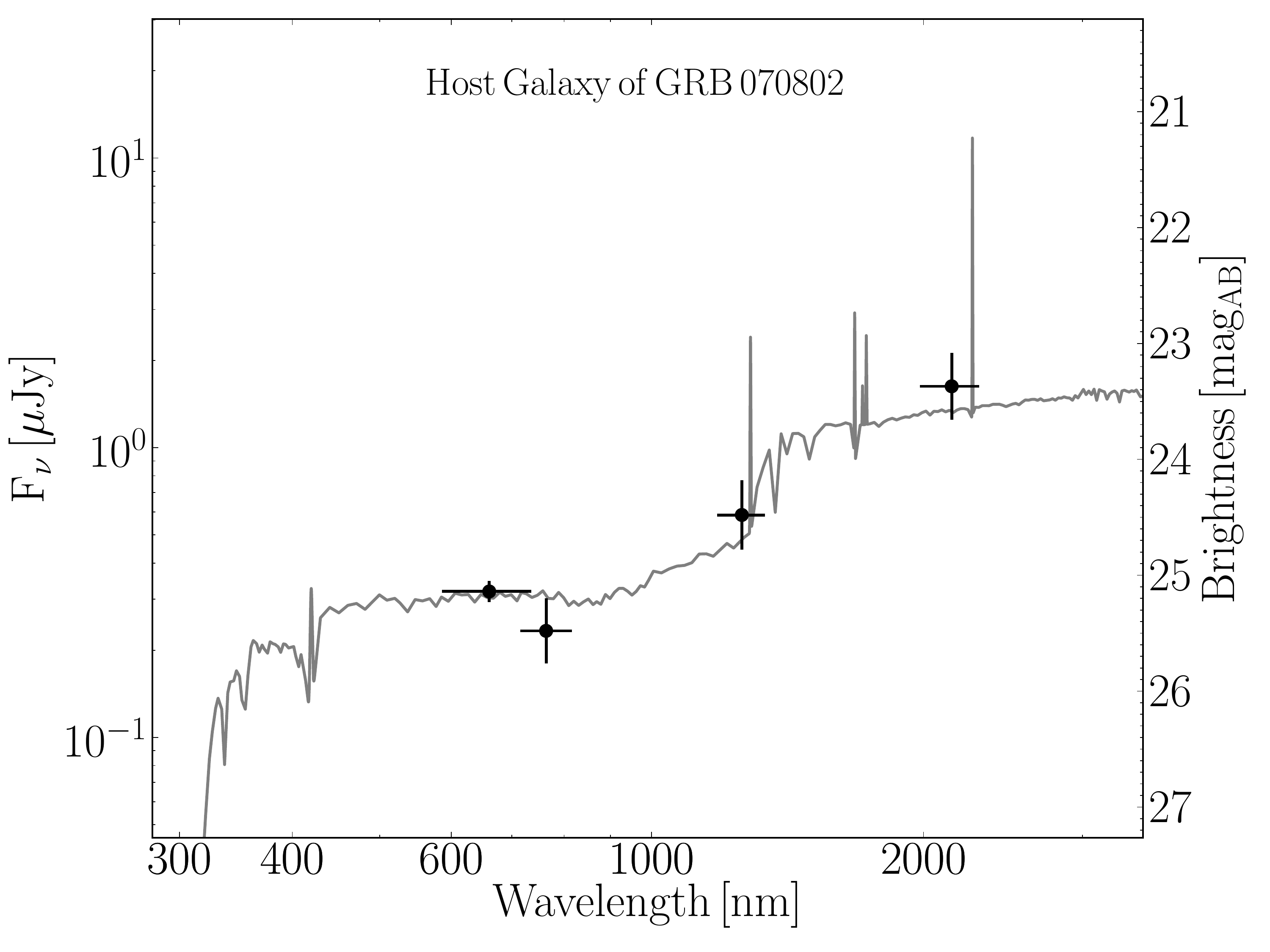}
\includegraphics[width=.9\columnwidth]{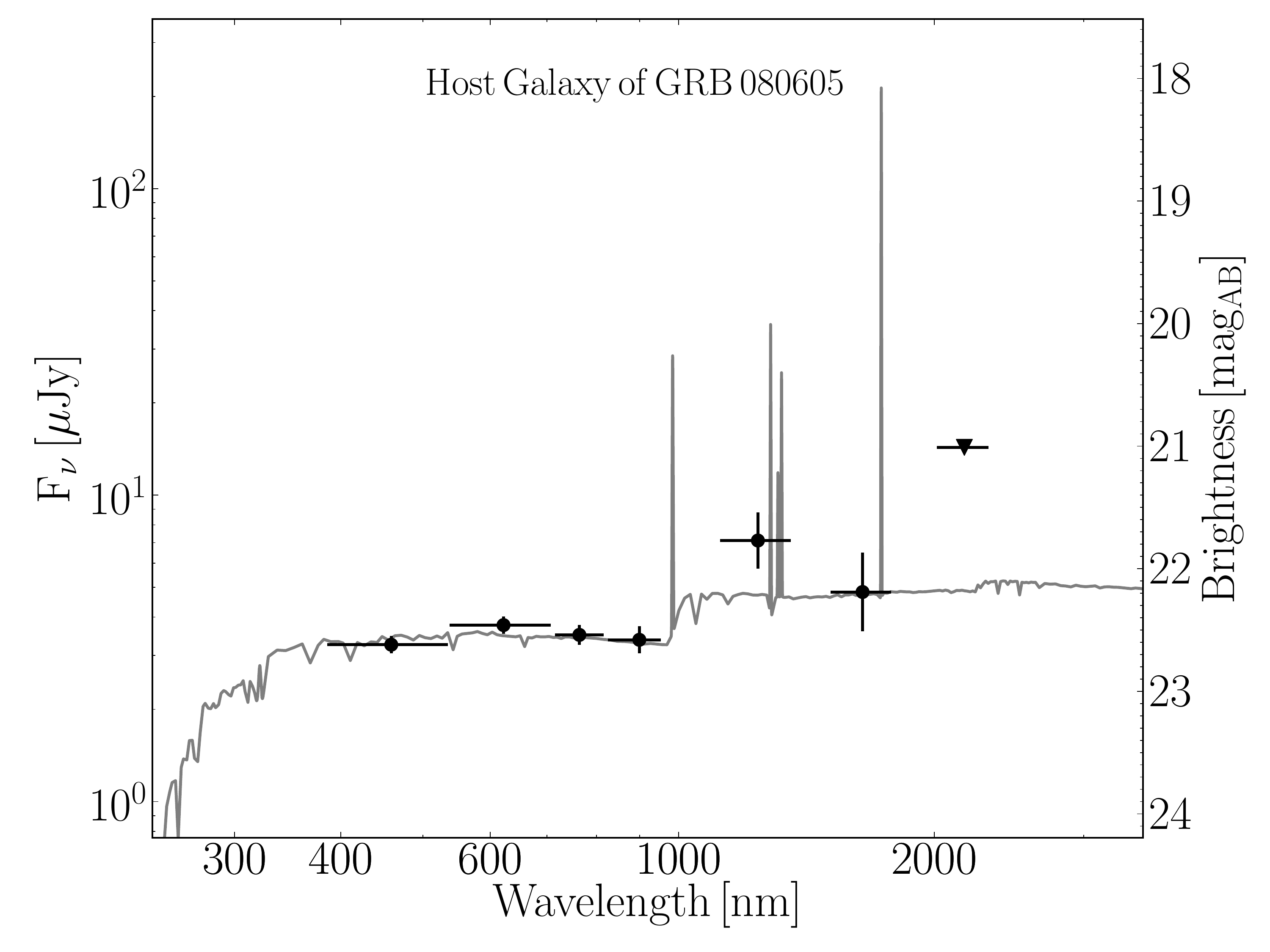}
\includegraphics[width=.9\columnwidth]{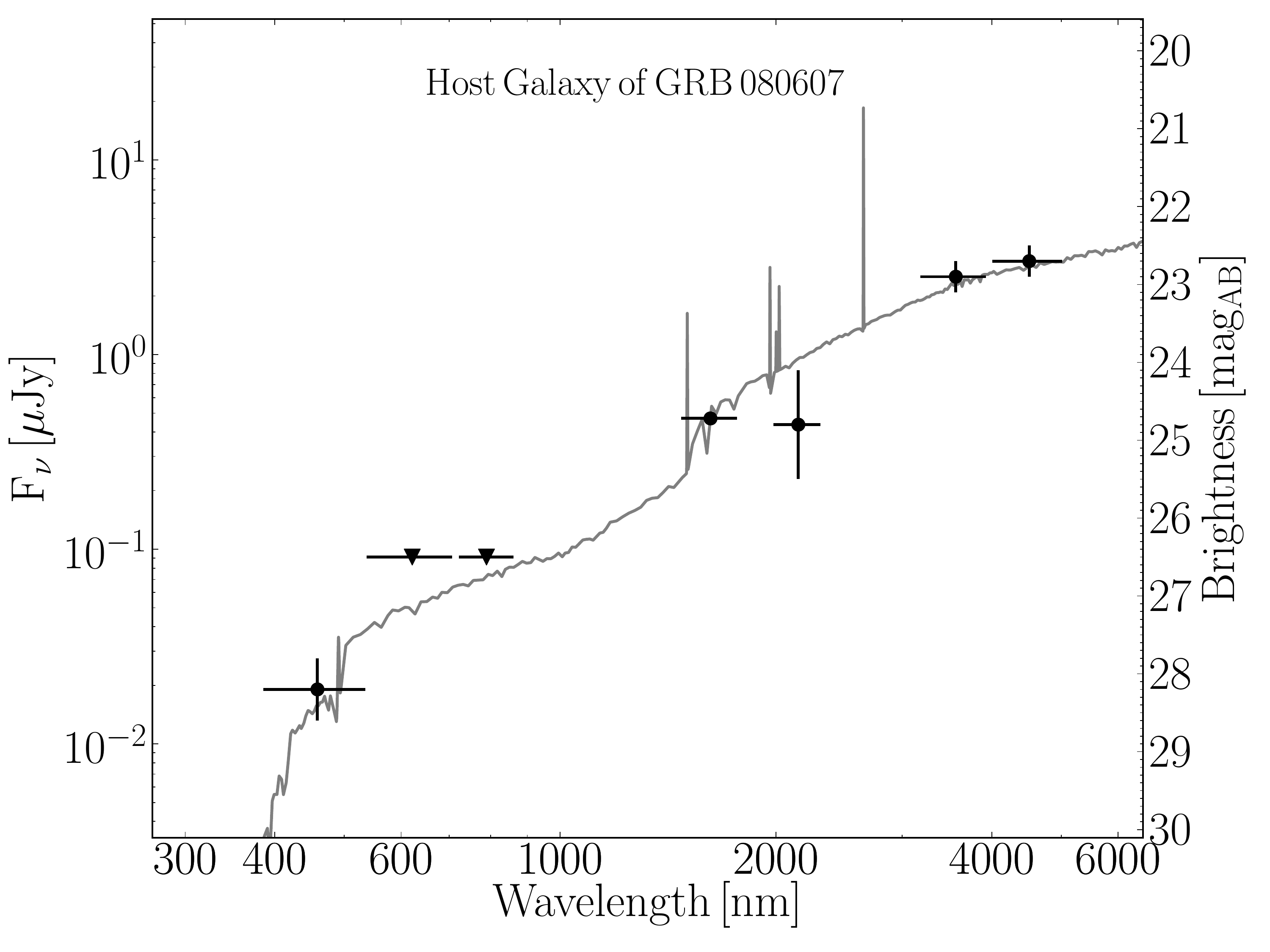}
\includegraphics[width=.9\columnwidth]{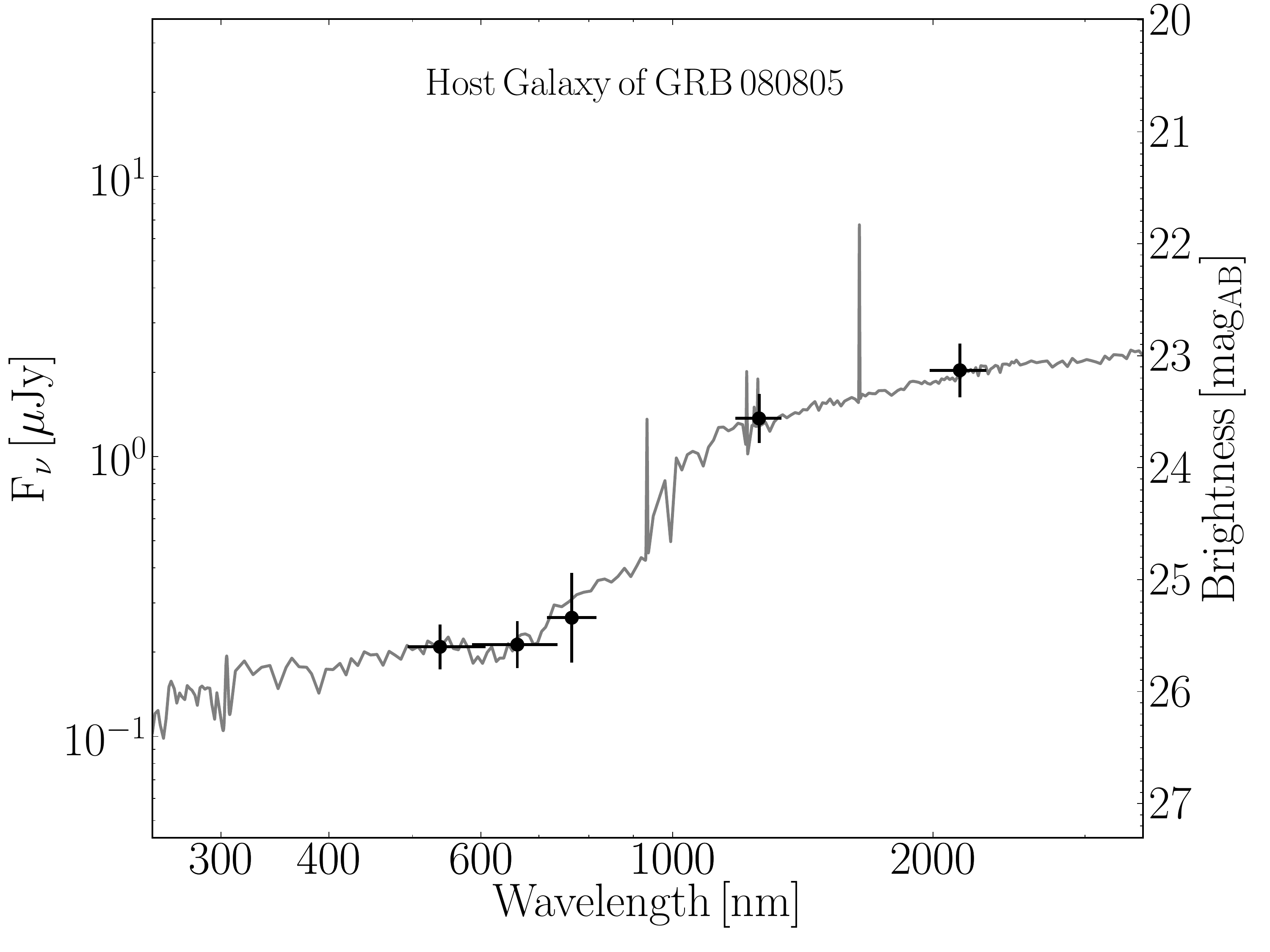}
\includegraphics[width=.9\columnwidth]{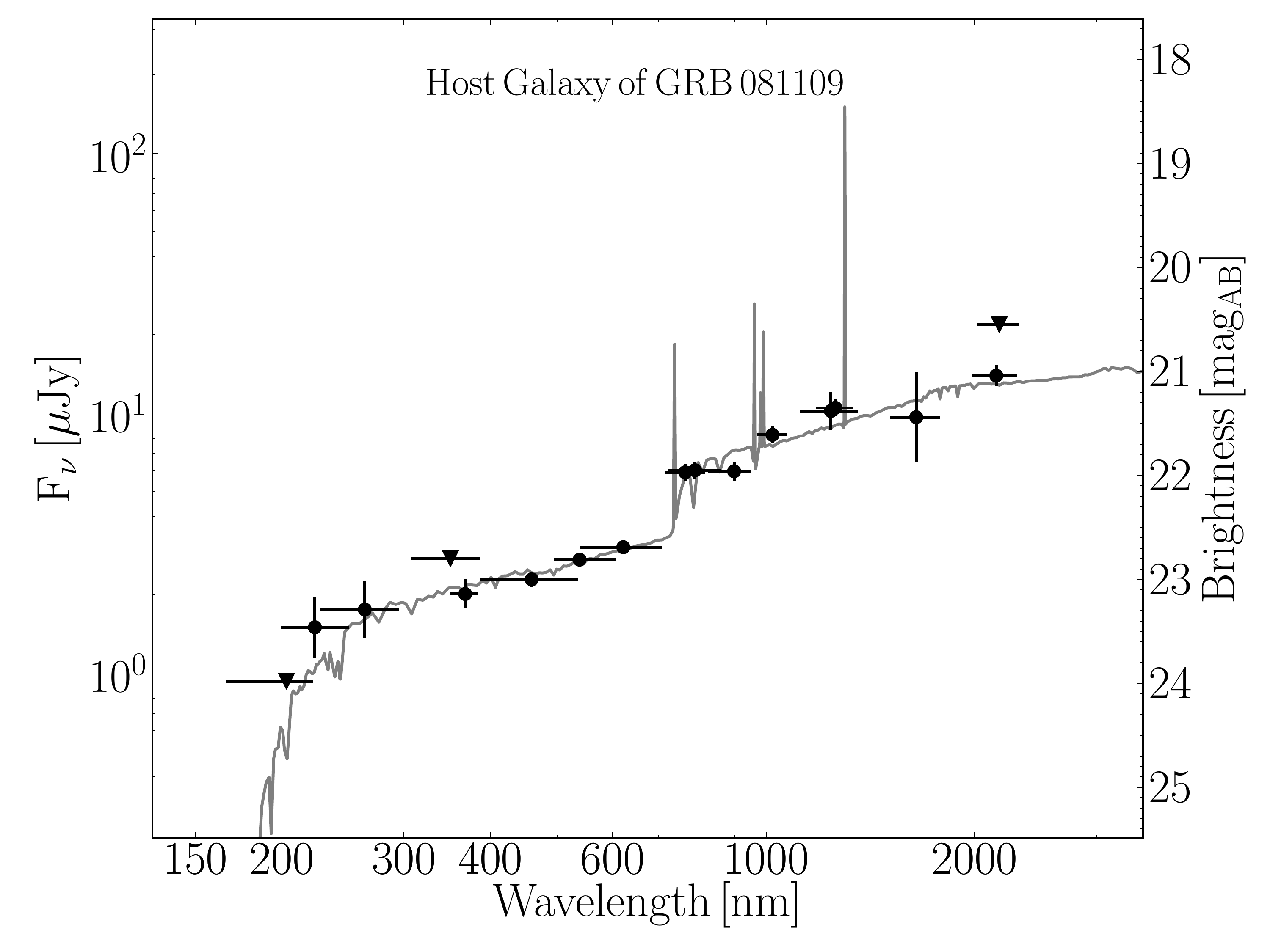}
\includegraphics[width=.9\columnwidth]{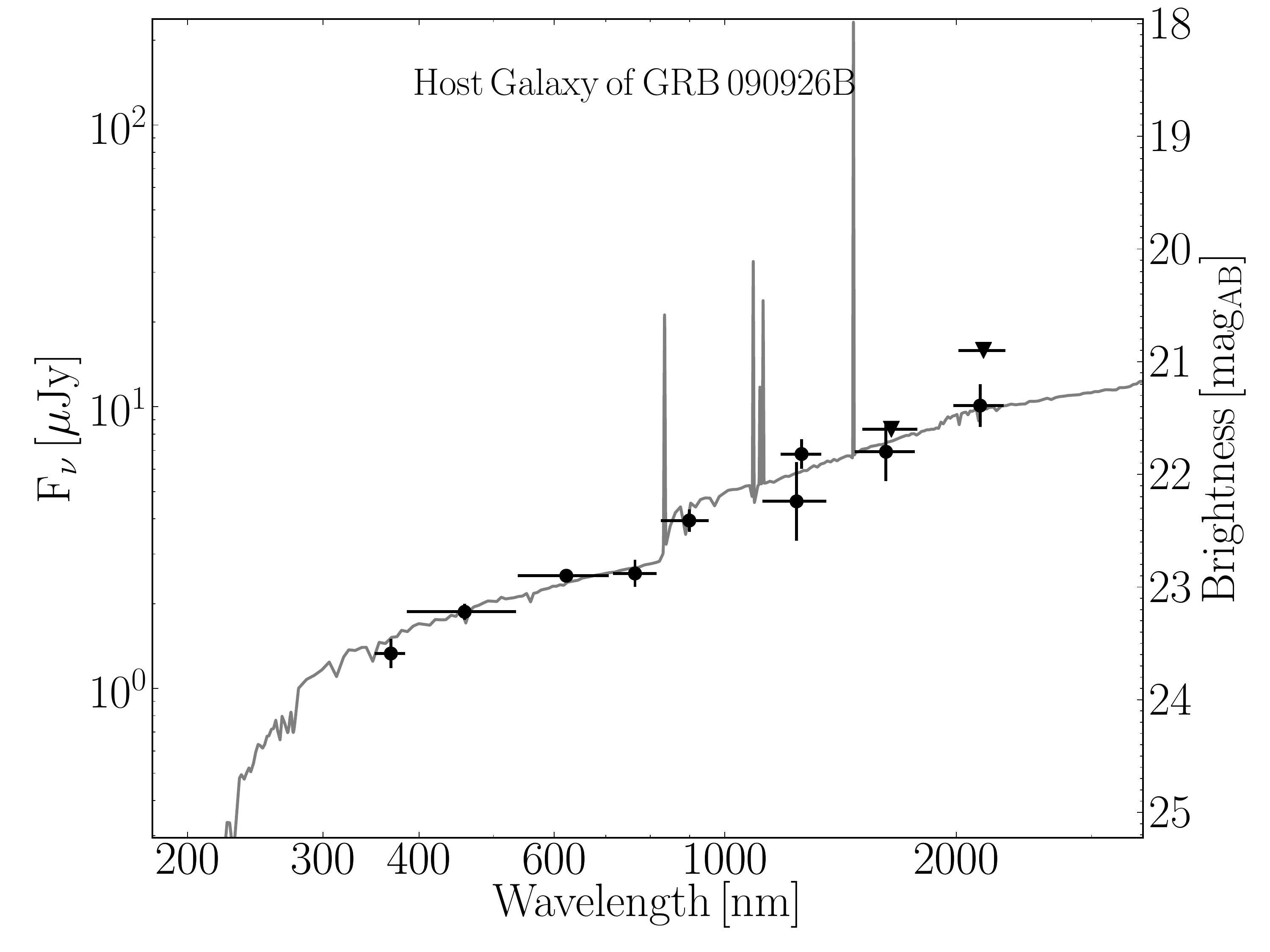}
\includegraphics[width=.9\columnwidth]{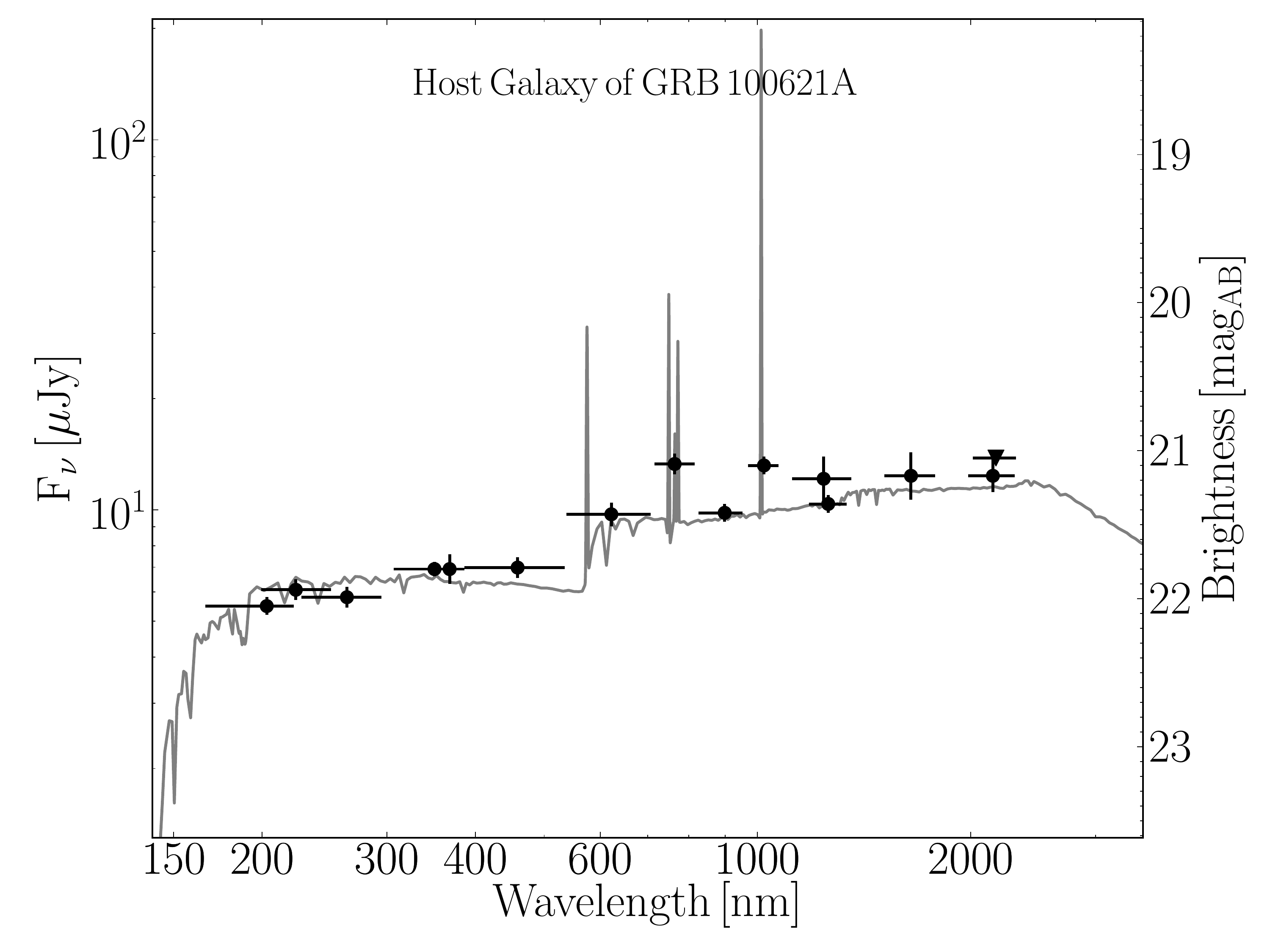}
\caption{SEDs of the host of in this sample and the best-fit galaxy model (solid line) in the observer's frame. Filled black circles represent photometric measurements, while downward triangles denote 3$\sigma$ upper limits. 
}
\label{hostfigs}
\end{figure*}

The galaxy hosting the strongly extinguished GRB~070306 at $z=1.496$ was previously discussed in \citet{2008ApJ...681..453J}. In addition to the public VLT imaging data (FORS $R$, ISAAC $J_sHK_s$), the host of GRB~070306 was observed with GROND (\gK simultaneously), and its SED (Fig.~\ref{hostfigs}) is further complemented by published $u$ and $I$-band data \citep{2008ApJ...681..453J}. The host is bright ($r^\prime = 23.1\,\rm{mag}$), mildly red\footnote{We compare GRB host colors against the median $(R-K)_{\rm AB} = 0.8\,\rm{mag}$ color from the SGL09 sample} with $(R-K)_{\rm AB}\sim1.5\,\rm{mag}$ and shows evidence of a $4000\,\AA$~break. The data are well fit with a non-extinguished ($A'_V < 0.6$)\footnote{For galaxies, we use $A'_V$, or $E'_{B-V}$ to indicate measured attenuation and effective reddening, since these quantities depend for example on the topology of the ISM and galaxy geometry \citep{2004ApJ...617.1022P}, and on galaxy scales are different from the corresponding values of a given extinction law.} host template, and yield an absolute AB magnitude of $M_B=-22.4\pm0.1\,\rm{mag}$, which, at $z\sim1.5$ corresponds to $\sim 1.7\, L^*$. The stellar mass of $\log (M_\ast [\Msun])= 10.4\pm0.2$ puts the galaxy among the most massive hosts compared to the sample of SGL09. The star-formation rate estimate from the rest-frame UV flux is $13_{-4}^{+11}$ \Msun/yr, which gives a specific star formation rate (SSFR = SFR/$M_\ast$) of $\sim 0.5$ Gyr$^{-1}$, or growth timescale (i.e., 1/SSFR) of 2 Gyr. The SFR is in reasonable agreement with the one derived from the [OII] emission line \citep{2008ApJ...681..453J}. We note, that the ISAAC $H$-band host image was obtained only 2.5~days after the GRB, and hence very likely contains a significant fraction of afterglow light, explaining the blue $H-K$ color. The physical properties of the host, however, are comparable if fit with or without the ISAAC $H$ filter.

\subsubsection{The host of GRB 070802}

The host of GRB~070802 at $z=2.452$ was discovered in deep FORS $R$ and ISAAC $K$ band imaging \citep{2009ApJ...697.1725E}. The afterglow SED is characterized by a prominent 2175~\AA~dust feature, and significant dust in the range of $A_V\sim 1\,\rm{mag}$ \citep{2008ApJ...685..376K, 2009ApJ...697.1725E}. To construct the optical/NIR host SED, the $R\sim25.2\,\rm{mag}$ host is further observed with EFOSC/NTT in $i$ and HAWK-I/VLT in the $J$-band. The galaxy is moderately red ($(R-K)_{\rm AB}\sim1.8\,\rm{mag}$) and its SED (Fig.~\ref{hostfigs}) shows a $4000~\AA$~break, but the age of the dominant stellar population is not well constrained by the available data ($\lesssim$~1~Gyr). There is no strong evidence for internal reddening, and the best-fit absolute magnitude is $M_B=-21.4\pm0.2\,\rm{mag}$, which is $\sim 0.6\, L^*$ at $z\sim2.5$ with $\log (M_\ast [\Msun]) = 9.7^{+0.2}_{-0.3}$. Using the rest-frame UV flux derived from the galaxy model fitting, an estimate for the SFR of $10_{-7}^{+30}$ \Msun/yr, and the SSFR of $\sim 2$ Gyr$^{-1}$ can be derived.

\subsubsection{The host of GRB 080605}

The host of GRB~080605 at $z=1.640$ was discovered in late GROND follow-up observations of the burst field 22 days after the GRB trigger. The afterglow SED shows evidence for a 2175~\AA~dust feature, and significant $A_V$ in the range of $\sim 0.5-1.3\,\rm{mag}$ \citep{2011A&A...526A..30G, 2011arXiv1102.1469Z}, and will be further discussed in Nicuesa et al. (in preparation). The host is bright ($r^\prime\sim 22.8\,\rm{mag}$), and blue with a flat $g^\prime-z^\prime$ color, and $(R-K)_{\rm AB} \sim 0.5\,\rm{mag}$ as estimated from the best fit galaxy model (Fig.~\ref{hostfigs}). The SED fit further yields $M_B=-22.6\pm0.2\,\rm{mag}$, which is $\sim 2.1\, L^*$ at $z\sim1.5$, and $\log (M_\ast [\Msun]) = 9.6^{+0.3}_{-0.2}$. The dominant stellar population of the host is young ($\tau = 0.06^{+0.18}_{-0.03}$~Gyr), and there is no evidence for reddening at the $2\sigma$ level ($E'_{B-V} \lesssim 0.22\,\rm{mag}$). The host galaxy is vigorously star-forming with a SFR of $40_{-20}^{+40}$ \Msun/yr and a SSFR of $\sim 10$ Gyr$^{-1}$.

\subsubsection{The host of GRB 080607}

The afterglow of GRB~080607 is heavily reddened ($A_V\sim 3\,\rm{mag}$), has a modest 2175~\AA~dust feature and is characterized by a strong neutral hydrogen absorber with roughly solar metallicity and molecular gas \citep{2009ApJ...691L..27P,  2010arXiv1009.0004P}. Data from \citet{2011ApJ...727L..53C} show a $R\sim27\,\rm{mag}$, very red host with a synthetic color of $(R-K)_{\rm AB} \sim 3\,\rm{mag}$ (Fig.~\ref{hostfigs}). The host is well described with an extinguished galaxy template ($A'_V \sim 1 - 2\,\rm{mag}$), and physical parameters of $M_B=-21.1\pm0.3\,\rm{mag}$, which is $\sim 0.3\, L^*$ at $z\sim3$, $\log (M_\ast [\Msun]) = 9.9^{+0.4}_{-0.6}$, an extinction corrected SFR of $40_{-26}^{+60}$ \Msun/yr, and SSFR of $\sim 8$ Gyr$^{-1}$. These values are in good agreement with previously published ones \citep{2011ApJ...727L..53C}.

\subsubsection{The host of GRB 080805}

GRB~080805 had a very red afterglow, where both an SED and spectral analysis showed a large dust column ($A_V \sim 1.0 - 1.5\,\rm{mag}$), and evidence for a 2175~\AA~dust feature \citep{2009ApJS..185..526F, 2011A&A...526A..30G, 2011arXiv1102.1469Z}. Its host is discovered in late EFOSC/HAWK-I imaging in five filters ($VRiJK$, see Fig.~\ref{hostfigs}), is relatively bright ($R\sim25.5\,\rm{mag}$) and red ($(R-K)_{\rm AB} = 2.5\,\rm{mag}$), with best-fit physical parameters of $M_B=-20.4\pm0.2\,\rm{mag}$ ($\sim 0.3\, L^*$ at $z\sim1.5$) and $\log (M_\ast [\Msun]) = 9.7^{+0.2}_{-0.2}$. The remaining galaxy properties are not well constrained by the available data, yielding a limit on the galaxy reddening $E'_{B-V} \lesssim 0.7\,\rm{mag}$, an age of the stellar population of $0.53^{+0.76}_{-0.42}$~Gyr, a SFR of $6_{-5}^{+25}$ \Msun/yr, and SSFR of $\sim 1$ Gyr$^{-1}$.

At a distance of 2.5$\arcsec$ north-east of the afterglow/host position, there is another $R\sim25\,\rm{mag}$, and even redder ($(R-K)_{\rm AB} \sim 4\,\rm{mag}$) galaxy, a plausible candidate for the strong \ion{Mg}{II} absorbing system at $z=1.20$ reported in \citet{2009ApJS..185..526F}.

\subsubsection{The host of GRB 081109}

\begin{figure}
\centering
\includegraphics[width=\columnwidth]{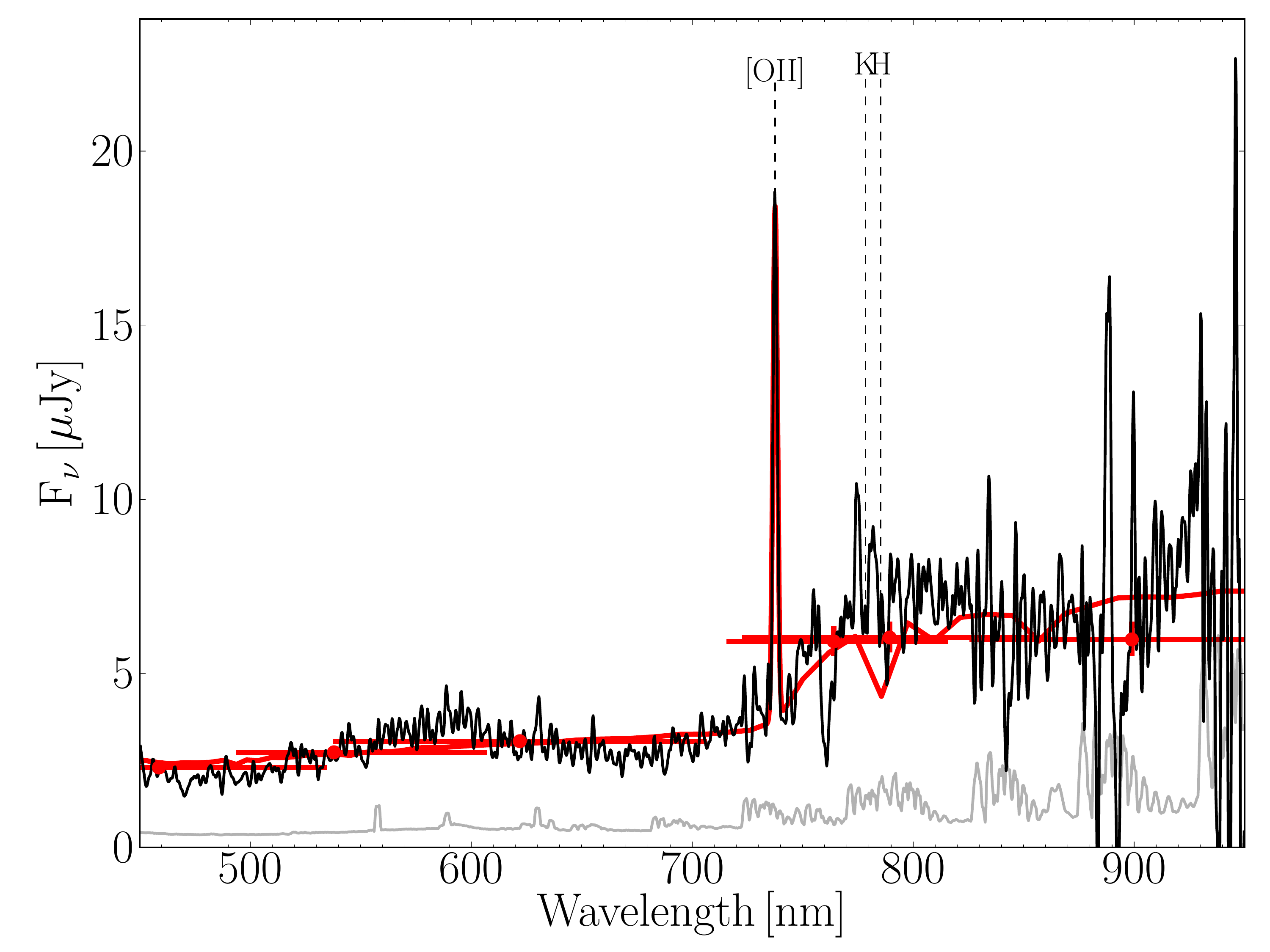}
\caption{Wavelength and flux-calibrated FORS2 300V spectrum of the host of GRB~081109 in black. The thin grey line shows the sky spectrum. The red line is the best-fit galaxy model obtained from the available photometry and red points are the photometric measurements.}
\label{081109spec}
\end{figure}

GRB~081109 is the only burst in the sample where no spectroscopic redshift was available in the literature. However, the host is bright ($r^\prime\sim 22.7\,\rm{mag}$), moderately red ($(R-K)_{\rm AB} = 1.6\,\rm{mag}$) and our host spectrum (see Fig.~\ref{081109spec}) reveals a single emission line above a well-detected continuum. Within the wavelength coverage of the spectrum ($\sim 370-950$~nm), this emission line is interpreted as [OII][$\lambda 3727$] at $z=0.9787 \pm 0.0005$. If it were any of the other prominent nebular lines (H$\beta$, [OIII], H$\alpha$), we would have expected to detect [OII][$\lambda 3727$] in the spectrum as well. At this redshift, there is further spectroscopic evidence for a Balmer break, and a tentative absorption of the \ion{Ca}{II} HK doublet (see Fig.~\ref{081109spec}).

The host SED is shown in Fig.~\ref{hostfigs} and well fit with a young ($\tau = 0.18^{+0.19}_{-0.07}$~Gyr) and reddened ($A'_{V} = 1.0\pm0.2\,\rm{mag}$) stellar population. The absolute magnitude of $M_B=-21.27\pm0.09\,\rm{mag}$ corresponds to $\sim 0.9\, L^\ast$ at $z\sim1$. The stellar mass obtained from the SPS fit is $\log (M_\ast [\Msun]) = 9.80^{+0.09}_{-0.09}$, with a SFR of $33_{-13}^{+19}$ \Msun/yr, which together yields a SSFR of $\sim 5$ Gyr$^{-1}$. The emission line flux of [OII][$\lambda 3727$] is $(1.8\pm0.2)\times10^{-16} \rm{erg/cm^2/s}$, which includes a systematic error contribution from the absolute flux normalization. This corresponds to a dust un-corrected star-formation rate of $13\pm4$ \Msun/yr \citep{1998ARA&A..36..189K}, or $48_{-16}^{+18}$ \Msun/yr when using the upper $A'_{V}$ measurement, which is in good agreement with the SFR from the UV continuum.

\subsubsection{The host of GRB 090926B}

Promptly after the trigger, \citet{2009GCN..9947....1F} reported the detection of the host galaxy of GRB~090926B based on a prominent [OII] emission line in a VLT/FORS spectrum. The host is also imaged in later GROND and NTT observations with a brightness of $r^\prime~\sim 23.0\,\rm{mag}$, and a mildly red color ($(R-K)_{\rm AB} \sim 1.5\,\rm{mag}$) as shown in Fig.~\ref{hostfigs}. The host SED is well fit with an extinguished host model ($A'_V = 1.4^{+0.3}_{-0.2}\,\rm{mag}$) with $M_B=-21.5\pm0.2$ ($\sim 0.9\, L^\ast$ at $z\sim 1.3$) and $\log (M_\ast [\Msun]) = 10.1^{+0.6}_{-0.5}$. The SFR based on the extinction corrected UV flux is $80_{-50}^{+110}$ \Msun/yr which is among the highest ever measured from optical data, but not well constrained given the uncertainty on the dust extinction properties. The constraints on the age of the stellar population and SSFR are weak, with values of $\tau=0.14^{+0.60}_{-0.09}$ Gyr and the SSFR of $\sim 7$ Gyr$^{-1}$, respectively.

\subsubsection{The host of GRB 100621A}

The host of GRB~100621A was reported very early after the trigger as a DSS2 source, providing a constant contribution to the afterglow in the UV/blue light curve \citep{2010GCN.10874....1U, 2010GCN.10878....1O}. In fact, the redshift of $z=0.542$ of GRB~100621A is based on emission lines from a bright host \citep{2010GCN.10876....1M}. The SED of the $r^\prime\sim 21.5$~mag host is well-sampled from the UV to the NIR and shown in Fig.~\ref{hostfigs}. In strong contrast to the extremely red afterglow $(R-K)_{\rm AB} \sim 5.8$~mag, the host is blue with an inferred color of $(R-K)_{\rm AB} \sim0.3\,\rm{mag}$ and $(uvw2-K)_{\rm AB} \sim0.9\,\rm{mag}$. The SPS host fit returns an intrinsic extinction of $A'_V=0.6_{-0.2}^{+0.1}\,\rm{mag}$ for a very young stellar population of age $\tau=0.05^{+0.07}_{-0.03}$ Gyr. This host has the lowest stellar mass in the presented sample with $\log (M_\ast [\Msun]) = 8.98^{+0.14}_{-0.10}$, and an absolute magnitude of $M_B=-20.68\pm0.08\,\rm{mag}$, which is $\sim 0.6\, L^\ast$ at $z\sim 0.5$. The SFR and SSFR are $13_{-5}^{+6}$ \Msun/yr and $\sim 14$ Gyr$^{-1}$, respectively.

\section{Discussion}

\subsection{Dust reddening in GRB afterglow SEDs}

The visual extinction measured from X-ray-to-NIR SED-fitting towards GRBs~070306 ($A_V^{\rm GRB} = 5.5^{+1.2}_{-1.0}\,\rm{mag}$), 081109  ($A_V^{\rm GRB} = 3.4^{+0.4}_{-0.3}\,\rm{mag}$) and 100621A ($A_V^{\rm GRB} = 3.8\pm0.2\,\rm{mag}$) are among the largest ever derived from optical/NIR data for GRB afterglows \citep[e.g.,][]{2003ApJ...585..638S, 2006ApJ...641..993K, 2011A&A...526A..30G}. They clearly show that a large dust column can be the dominating feature in a GRB afterglow SED. The dust properties inferred from afterglow measurements are well represented with local models in the rest-frame 300-1100~nm, and at the resolution of broad-band imaging \citep[see also e.g.,][]{2001A&A...369..373F,  2006ApJ...652.1011W, 2006ApJ...641..993K, 2007MNRAS.377..273S, 2007ApJ...661..787S}, while noteworthy exceptions exist in the literature \citep[e.g.,][]{2004ApJ...614..293S, 2010MNRAS.406.2473P, 2011arXiv1103.6130C}. The good fit provided by local dust extinction laws further suggests an abundance of small dust grains comparable to the MW/LMC or SMC. There is hence no direct evidence that the dust towards these GRBs through their hosts is different than observed along local sight-lines. A different dust grain size distribution would have been expected if the dust were located in the immediate vicinity ($R \lesssim 10^{19}$~cm) of the GRB and shaped through its intense radiation, i.e., through dust destruction \citep{2000ApJ...537..796W, 2001ApJ...563..597F, 2002ApJ...569..780D}. In addition, the metals-to-dust ratios for these afterglows are only a few times the Galactic value of $N_{H}/A_V\approx 2\times10^{21}$cm$^{-2}$/mag \citep{1995A&A...293..889P}. For un-extinguished GRB sight-lines this ratio is generally found to be a factor of 10-100 times larger than those of the Magellanic Clouds or Milky Way \citep[e.g.,][]{2001ApJ...549L.209G, 2004ApJ...608..846S}. This is suggestive of a dependence of the metals-to-dust ratio on the amount of visual extinction. We will return to this issue in Sect.~\ref{avnhcon}.

\subsection{The hosts of the dustiest afterglows}
\label{dustyhosts}
The general properties of the selected GRB host galaxies are diverse. They have $(R-K)_{\rm AB}$ colors ranging from flat and blue $(R-K)_{\rm AB}\sim 0 \,\rm{mag}$ to extremely red $(R-K)_{\rm AB}\sim3\,\rm{mag}$ with an average color of $\langle(R-K)_{\rm AB} \rangle = 1.6\,\rm{mag}$, and host extinction values between $A'_V \sim 0\,\rm{mag}$ and $A'_V \sim 2\,\rm{mag}$. Also, their stellar mass and absolute magnitude distributions are broad, with values between $\log (M_\ast [\Msun]) = 9.0$ to $\log (M_\ast [\Msun])=10.4$ ($\langle \log (M_\ast [\Msun])\rangle = 9.8\pm0.4$) and $M_B$ between $-20.3$~mag and $-22.6$~mag ($\langle M_B \rangle = -21.3\pm0.6\,\rm{mag}$). These absolute brightnesses are in a range between several tenths to few $L^\ast$ ($\langle L \rangle = 0.9L^\ast$) as compared to the general field galaxy population at the same redshift. The average SFR and SSFR are about $30$ \Msun/yr and $\langle \log \rm{SSFR}\, \rm{[yr]}\rangle \sim -8.3$, respectively. The average growth time is $\sim0.2$ Gyr, which illustrates that not only optically-selected hosts, but also those of highly-reddened afterglows are very efficient in producing stars. A rough estimate on the metallicity of the hosts can be obtained if these GRB hosts follow the fundamental plane as defined from nearby SDSS galaxies \citep{2010MNRAS.408.2115M}. With a given stellar mass and SFR, the host galaxies in this sample are expected to have metallicities in a range between $12+\log(\rm{O/H})\sim8.2$ and  $12+\log(\rm{O/H})\sim8.9$, with an average of $12+\log(\rm{O/H})\sim8.6$. We caution that the SFRs were derived using the rest-frame UV flux, which is quite sensitive to the dust extinction properties.

Although not well-constrained in all cases, the average luminosity-weighted effective reddening inferred from host photometry is typically smaller or equal to that measured from afterglow observations. This is not a particularly surprising result, given that the sample selection was based on high visual extinctions of the afterglow SEDs in the first place. It directly indicates some variation in the dust distribution of the hosts, which again is not a surprising result, given the differences in extinction properties along different sight-lines through the diffuse ISM to Giant Molecular Clouds in the Local Group \citep[e.g.,][]{2003ApJ...594..279G, 2004ASPC..309...33F}, and the geometrical differences between a single sight-line and an extended distribution of star-light and dust \citep[e.g.,][]{1997ApJ...487..625G, 1998ApJ...509..103S}.

One intriguing case is the host of GRB~100621A. Although having one of the most extinguished afterglows ever detected (even in the presented sample), its host shows very blue colors, and is one of the youngest and the least massive galaxy in this work. This particular example provides evidence for a patchy dust component where the geometry of the dust distribution and not the properties of the host galaxy makes the single GRB sight-line dust-enriched.

\subsection{Comparison to previous GRB host samples}

One key result of this study is the success rate of the discovery of the selected hosts. Out of eight hosts, which were selected given their afterglow properties (hence a selection independent on host properties, in particular galaxy brightness), all are luminous enough to detect in optical ground-based imaging. This fraction is significantly larger than expected from a host sample based on XRT detections \citep{2008mgm..conf..726F, 2009AIPC.1111..513M}. The effect of an increased detection rate is even stronger in the NIR: Seven out of eight are detectable in the $K$-band, while this fraction is only $\sim$35\% for the general host population of \textit{Swift}/GRBs \citep{2009AIPC.1111..513M}. Partially, this is the result of the lower average redshift of the selected hosts ($\langle z_{A_V} \rangle = 1.5$) as compared to all \textit{Swift} GRBs with $\langle z_{Swift} \rangle \sim 1.9$.

\begin{figure}
\centering
\includegraphics[width=\columnwidth]{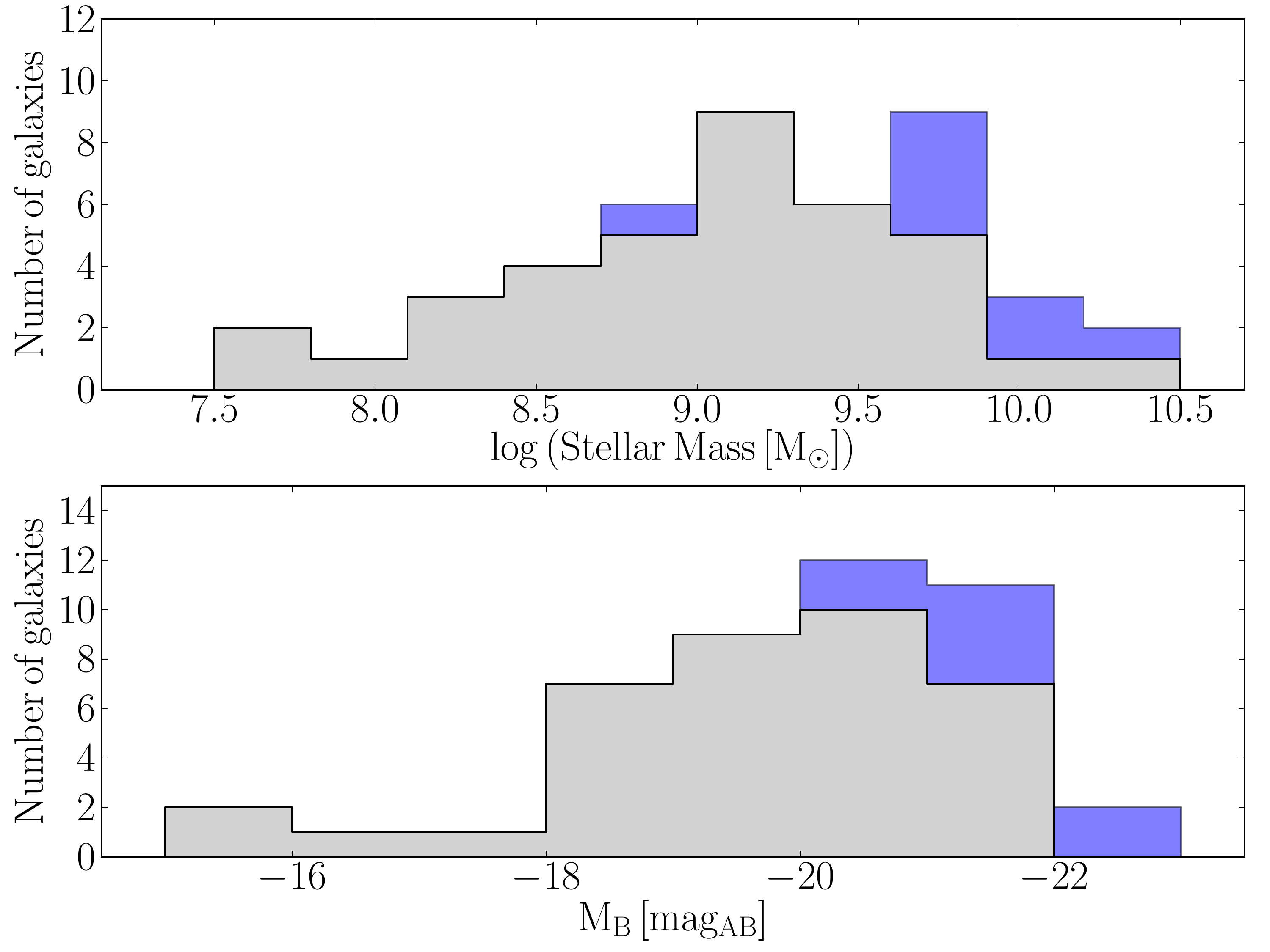}
\caption{Distribution of stellar masses and luminosities of the hosts of highly extinguished afterglows (blue) and the host sample from SGL09 (grey).}
\label{masshist}
\end{figure}

The lower redshift is however not the only reason for the high detection rate. The selected hosts are on average redder and, as shown in Fig.~\ref{masshist}, have typically higher luminosities and stellar masses than the (sensitivity-limited) SGL09 sample which has $\langle(R-K)_{\rm AB} \rangle = 0.8\,\rm{mag}$, $\langle M_B \rangle = -19.6\pm1.5\,\rm{mag}$ and  $\langle \log M_\ast [\Msun]\rangle = 9.1 \pm 0.6$. A two-sample K.-S. test returns $p$-values of 0.002 for the stellar mass, and 0.006 for the absolute magnitude distributions respectively, which is tentative evidence that both distributions are not drawn from the same parent sample. However, given the small sample size of only eight high-$A_V$ events, larger samples are required to statistically establish the existence of a difference at higher significance. Of course, both distributions are drawn from the same physical parent sample (GRB hosts), indicating that the different selection criteria probe different host properties.

\begin{figure}
\centering
\includegraphics[width=\columnwidth]{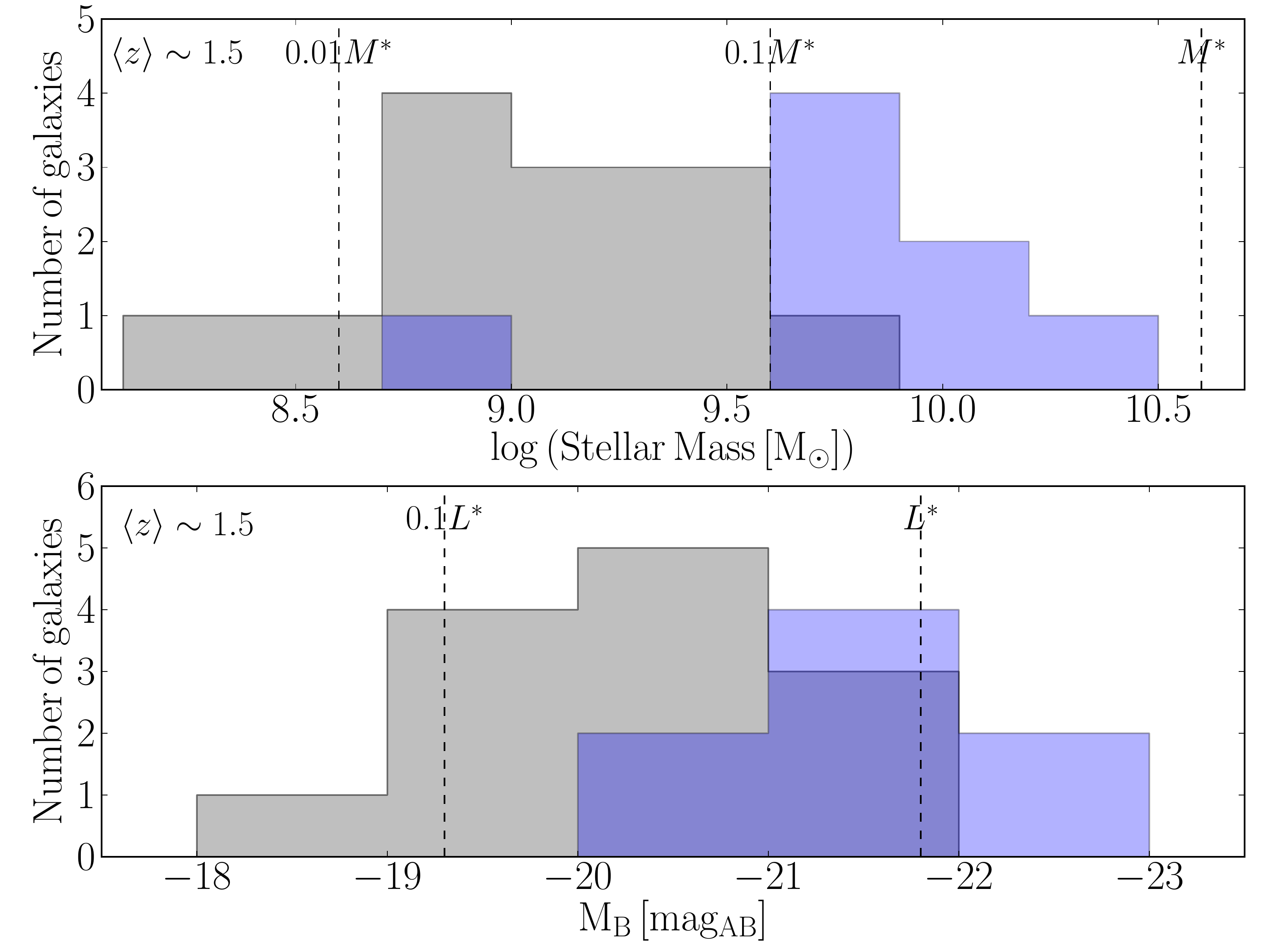}
\caption{Distribution of stellar masses and luminosities of the hosts of highly extinguished afterglows (blue) and a subsample of SGL09 (grey) with $\langle z \rangle \sim 1.5$.}
\label{masshist2}
\end{figure}

A possible explanation of the different host properties would be the now on-average higher redshift relative to the SGL09 sample, where star-formation was driven by more massive galaxies as compared to the more nearby Universe \citep[e.g.,][]{1996AJ....112..839C, 2004ApJ...615..209H}. To test this hypothesis, we selected a sub-sample from SGL09 with a median redshift comparable to the hosts in this work. This essentially removes all $z < 1$ SGL09 hosts and leaves only 13 events for comparison (see histograms in Fig.~\ref{masshist2}). Despite the small number statistics, the $M_\ast$ and $M_B$ values are again placed at the high-mass and high-luminosity end of their respective distribution, and a K.-S. test is also marginally suggestive of a difference ($p$-values of 0.001 and 0.034 for the masses and absolute magnitudes). 

We conclude that by selecting extinguished afterglows we are very likely probing a more luminous, massive and chemically-evolved population of GRB hosts. 

As it is clear that these were largely missing from previous samples due to their poor localizations, there is a selection bias and the host population is missing most of its massive, evolved and metal-rich members. As a direct consequence, GRB hosts trace the global SFR closer than indicated in studies which are based on host samples of optically selected GRB afterglows, and the apparant deficiency of high-mass host galaxies is at least partially a selection effect.

Similar conclusions apply for all galaxies hosting afterglows that show a significant 2175~\AA~dust feature in their SED. Four out of five currently known afterglows are within the presented sample, which argues for a direct connection between large dust columns and the presence of the UV~bump \citep[see e.g.,][]{2011A&A...526A..30G, 2011arXiv1102.1469Z}. 
Their, on average, more massive and luminous hosts suggest a qualitative relation between the stellar mass of a galaxy and the presence of a 2175~\AA~feature, where the latter is only present in fairly massive and metal-enriched galaxies \citep[see also e.g.,][]{2005A&A...444..137N}. Conversely, a strong 2175~\AA~feature in an afterglow SED is also very likely a good proxy for the stellar mass and luminosity of the GRB host.


\subsection{Metals-to-dust ratios in context}

\label{avnhcon}

The ratio between the line-of-sight extinction and the total metal column for GRB afterglows has been investigated in a number of papers \citep[e.g.,][]{2001ApJ...549L.209G, 2004ApJ...608..846S, 2004ApJ...614..293S, 2006ApJ...641..993K, 2007MNRAS.377..273S, 2010MNRAS.401.2773S, 2011A&A...526A..30G, 2011arXiv1102.1469Z} where ratios typically much higher than the ones observed in the Local Group were derived. Measurements for different Galactic sight-lines \citep[e.g.,][]{1995A&A...293..889P, 2009MNRAS.400.2050G} show an almost universal value of around $N_{H}/A_V \approx 2\times10^{21}$cm$^{-2}$/mag, while the matter probed by afterglows can yield metals-to-dust ratios up to and sometimes even above $100$-times higher \citep[e.g.,][]{2007ApJ...660L.101W, 2010ApJ...720..862R}.

\subsubsection{An anti-correlation between metals-to-dust ratio and sight-line extinction}

\begin{figure}
\centering
\includegraphics[width=\columnwidth]{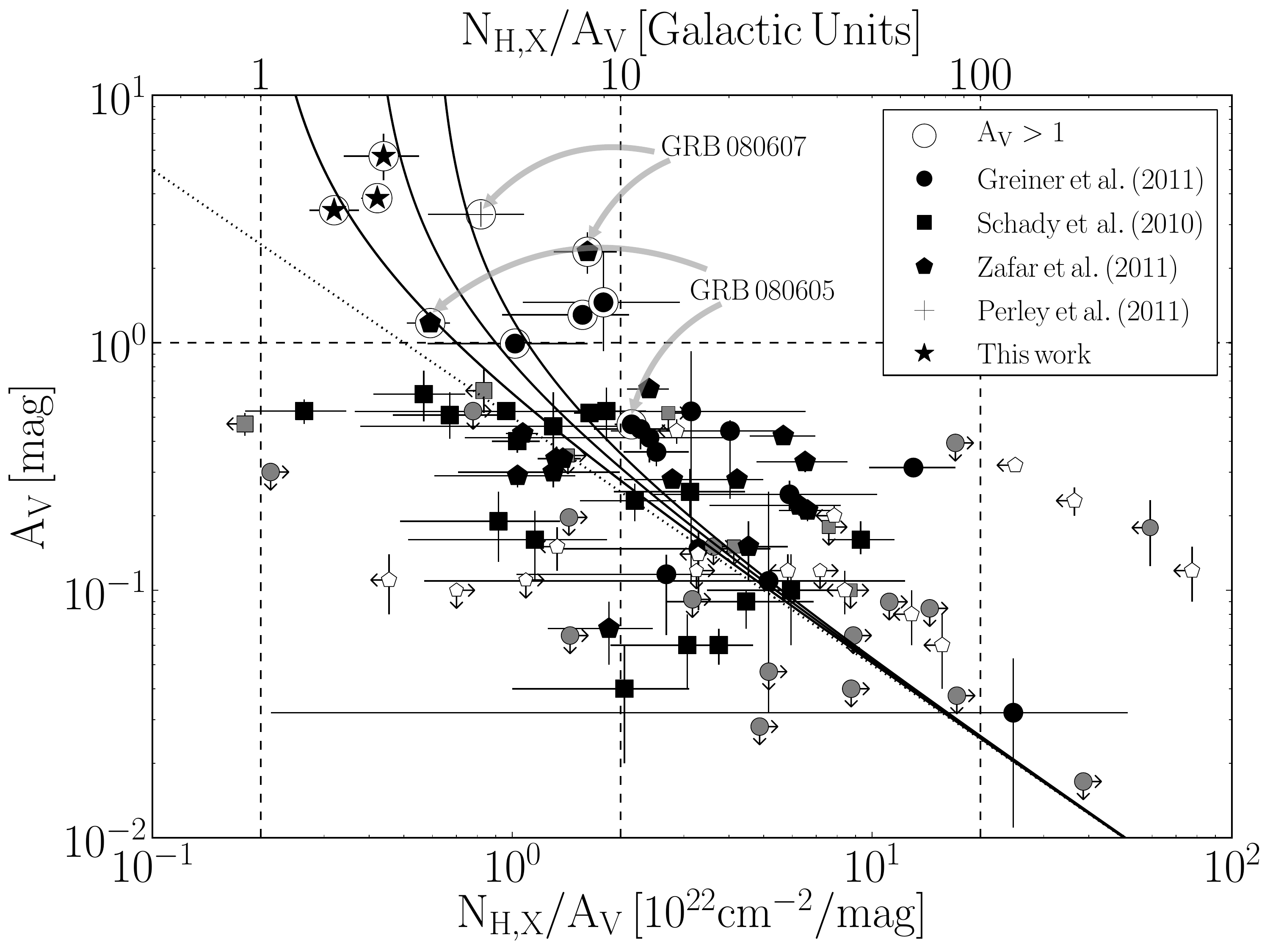}
\caption{Metals-to-dust ratio versus sight-line extinction for GRB afterglows. The horizontal dashed lines marks the selection criterion for GRBs to enter this sample. Vertical dashed lines illustrate metals-to-dust ratios of 1, 10 or 100 times the Galactic value. Solid lines show the toy model of two physically independent absorbers, where one is fully devoid of dust with a $H$-equivalent metal column of $N_{H, X} = 10^{21.5}$cm$^{-2}$ and represented by the dotted line, while the other is neutral and has metals-to-dust ratios of 1, 2 or 3 times the Local Group value. Two individual cases (GRBs~080605 and 080607) illustrate the scatter between the analysis of different data sets.}
\label{avnhratio}
\end{figure}

Figure~\ref{avnhratio} shows the $N_{H, X}/A_V^{\rm GRB}$ ratio for a large number of afterglows and illustrates its dependence on the sight-line dust extinction. With the afterglows in this work, there is now for the first time reasonable coverage in the $A_V^{\rm GRB} \sim 1-5\,\rm{mag}$ range. Intriguingly, the metals-to-dust ratio is strongly anti-correlated with $A_V^{\rm GRB}$, confirming the tentative trend reported by \citet{2009AJ....138.1690P}. A Spearman rank-order correlation analysis for the combined sample in Figure~\ref{avnhratio} returns a correlation coefficient $\rho = -0.63$, with a two-tailed $p$-value of $3\times10^{-7}$, in strong contrast to a constant, universal $N_{H, X}/A_V^{\rm GRB}$ ratio. 

There are two straight forward ways to reconcile this result: One is a dependence of the metals-to-dust ratio on the specific environment such that evolved and dust-enriched hosts are more efficient in forming dust out of their metals (and in fact we do observe on average larger stellar masses for the hosts of high $A_V^{\rm GRB}$ afterglows). The other is the presence of two physically independent absorbers, where the first dominates the total metal column, the other the visual extinction measurements. This trivially produces a non-correlation between $N_{H, X}$ and $A_V^{\rm GRB}$, and consequently an $N_{H, X}/A_V^{\rm GRB}$ to $A_V^{\rm GRB}$ anti-correlation.

Some outliers of the metals-to-dust anti-correlation (Fig.~\ref{avnhratio}) might be explained with difficulties of measuring the respective physical parameters. This is also illustrated by the example of two individual events (GRBs~080605 and 080607) where different values have been published in the literature. Assumptions on the continuum emission and the extinction law, notably the total-to-selective reddening $R_V$, but also its parametrization can affect the $A_V^{\rm GRB}$ measurement. In addition, there is the possibility that the cooling break is located close to or within the range of the UV/optical/NIR measurements. In a standard analysis, the introduced curvature caused by the spectral break is then interpreted as an increased dust column \citep{2011A&A...526A.153K}. The $N_{H, X}$ measurements are prone to errors as well: spectral variation intrinsic to the afterglow can lead to wrong estimates on the soft X-ray absorption \citep[e.g.,][]{2007ApJ...663..407B}.

\subsubsection{Metals-to-dust ratio compared to host mass}

\begin{figure}
\centering
\includegraphics[width=\columnwidth]{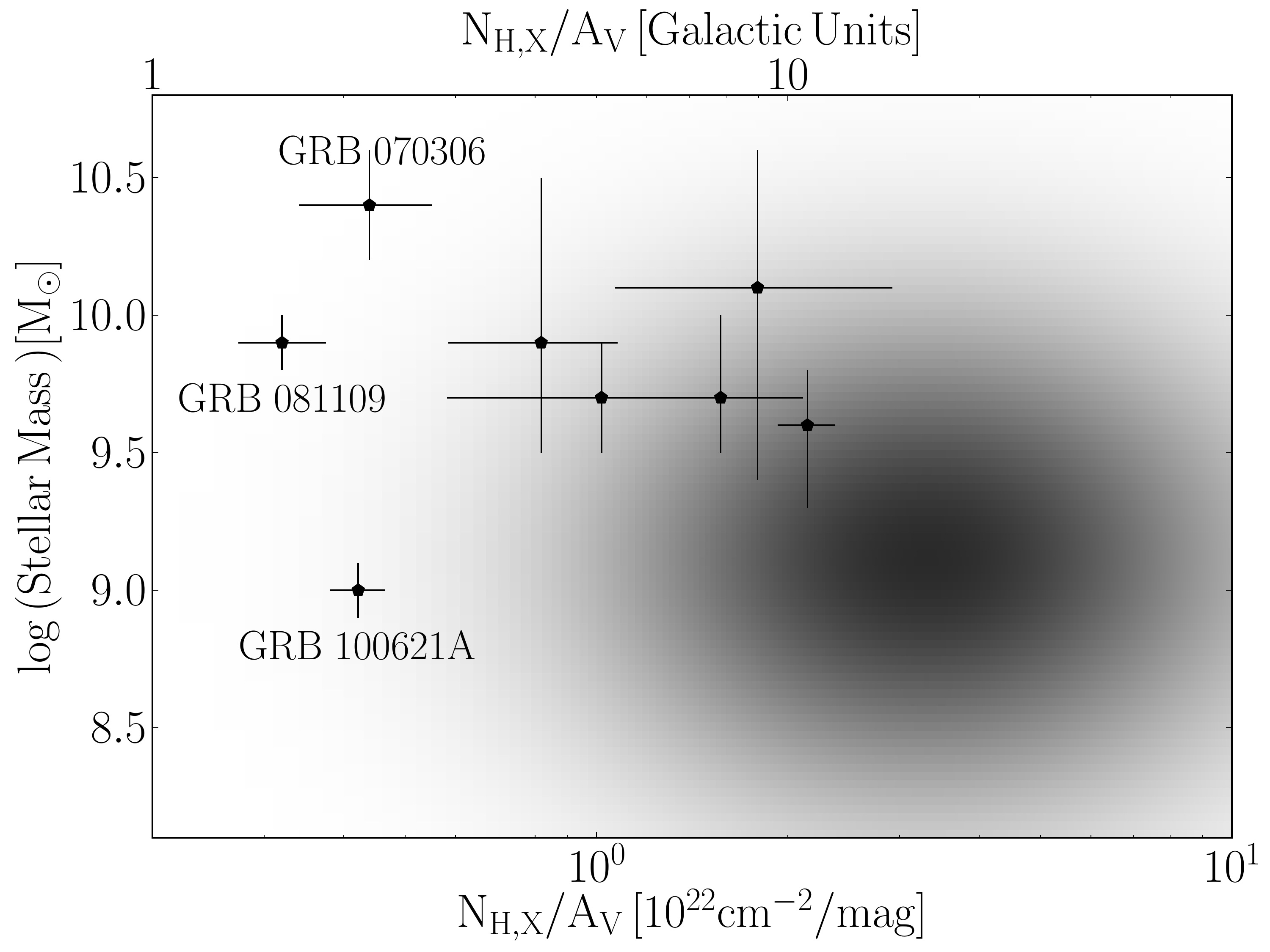}
\caption{Total metals-to-dust ratios for GRB afterglows versus stellar mass of their host galaxies. Black data are galaxies hosting a highly-extinguished afterglow using the first values of Table~\ref{tab:agres}, where the three hosts with the highest $A_V^{\rm GRB}$ are labeled. The shaded area indicates the probability distribution for optically-selected GRBs, represented by a log-normal distribution of $N_{H, X}/A_V^{\rm GRB}$ based on \citet{2010MNRAS.401.2773S} and a Gaussian distribution in $\log (M_* [M_{\sun}])$ based on SGL09.}
\label{avnhmass}
\end{figure}

As shown in Sect.~\ref{dustyhosts}, the hosts of dusty afterglows are on average more massive and luminous than their non-extinguished counterparts, but there is a broad range of galaxy properties and the only common feature between all afterglows/hosts in this work is hence the dusty line of sight. In particular, if the environment were responsible for the observed $N_{H, X}/A_V^{\rm GRB}$ to visual extinction anti-correlation, we would expect the metals-to-dust ratio for GRB~100621A to be comparable to the bulk of optically bright afterglows. It is, however, one of the lowest ever observed for GRB afterglows and a factor of 5 lower than the median for afterglows with hosts of similar mass (see Fig.~\ref{avnhmass}). Although we note that number counts are still too low to derive strong constraints with high statistical significance, this suggests that the specific host environment is not responsible for the observed dependence on the metals-to-dust ratio to the visual extinction.

\subsubsection{A system of two absorbers}

In the second scenario, two to first order physically independent (one neutral, host-galaxy related, one ionized, circumburst specific) columns of material contribute to the observed absorption, denoted as $N_{H, \rm{neutral}}$, $N_{H, \rm{ion}}$ in the following. $N_{H, X}$ measures the sum of both, whereas the $A_V^{\rm GRB}$ column would only be associated with the $N_{H, \rm{neutral}}$ absorber. Such a circumburst environment is not unexpected: The intense afterglow radiation should not only photo-ionize the vicinity of the burst, but also destroy the associated dust to large amounts \citep{2000ApJ...537..796W, 2002ApJ...569..780D, 2003ApJ...585..775P}, albeit with different effective radii.

For $A_V^{\rm GRB}\sim 0.1\,\rm{mag}$ sight-lines, $N_{H, \rm{neutral}} \ll N_{H, \rm{ion}}$ directly results in a large metals-to-dust ratio. With an increasing $A_V^{\rm GRB}$ column of around $1\,\rm{mag}$, the $N_{H, \rm{neutral}}$ absorber contributes significantly to the total metal column ($N_{H, \rm{neutral}} \approx N_{H, \rm{ion}}$): this is illustrated in Fig.~\ref{avnhratio} by the solid lines, where the $N_{H, X}/A_V^{\rm GRB}$ ratio asymptotically reaches 1, 2 or 3 times the Local Group value. Even larger $A_V^{\rm GRB}$ columns than present in this work can test this hypothesis. For an $A_V^{\rm GRB}\gtrsim10\,\rm{mag}$ sight-line, for example, $N_{H, \rm{neutral}}$ is much larger than $N_{H, \rm{ion}}$ and the expected metals-to-dust ratio would be comparable to its intrinsic value, and to the Local Group value (if universal).

For the bulk of standard, un- or mildly extinguished afterglows, the large column of ionized metals with an equivalent $N_{H, \rm{ion}} \sim 10^{21-22}$~cm$^{-2}$ in the circumburst material dominates the total $N_{H, X}$ measurement, whereas the visual extinction is very likely caused by dust further out, either in the diffuse ISM, or localized in interstellar clouds (see also Sect.~\ref{dustgeo}). Hence, for optically bright GRBs the total metal absorption as probed by the soft X-ray absorption is neither a good measure nor a direct tracer of the dust extinction along the line of sight. Or, in other words, an $N_{H, X}$ column even as large as $10^{22}$~cm$^{-2}$ does not necessarily imply a significant visual extinction \citep[see also, e.g.,][]{2001ApJ...549L.209G, 2006ApJ...641..993K, 2007MNRAS.377..273S}.

Also, the shape of the extinction law, which is typically consistent with sight-lines through galaxies of the Local Group argues against an association of the absorbing dust with the immediate vicinity of the burst. The steepness of the UV rise of the extinction law as probed by GRB afterglows is comparable to the one of the LMC \citep[][]{Schady2011}, generally suggestive of an abundance of small dust grains. These grains, however, are expected to be destroyed first and the grain-size distribution function would be skewed to larger grains. This would result in flat, or grey extinction properties, in contrast to the ultra-red afterglows of GRB~081109, or GRB~100621A, for example.

We thus conclude that the anti-correlation between metals-to-dust ratio and sight-line extinction indicates the presence of two absorbing systems, which are to first order physically independent. One of them is dusty, the other ionized and dust-free, where the former is probed by the optical/NIR data and the latter typically dominates the $N_{H, X}$ measurement.

\subsection{Location and geometry of the absorbing dust column}
\label{dustgeo}

A natural question about the nature of the absorbing dust, gas and metal columns detected in the afterglow SEDs and spectra is their locations, and if they are directly related to the burst environment. A number of previous studies have already revealed some clues about the geometry of the absorbing matter, which have been derived quite exclusively from sight-lines with low total dust content: The distance of the cold-neutral material, linked to the DLA and the low-ionization metal absorption lines has been constrained to few hundreds of pc to few kpc \citep{2006ApJ...648...95P, 2007A&A...468...83V}. In contrast, the high metal column densities as derived from soft X-ray absorption were associated with a fully ionized circumburst medium up to few to several tens of pc \citep{2007ApJ...660L.101W, 2011A&A...525A.113S}. For bursts with largely unextinguished afterglows, most of the metals along the sight-line are typically in a highly ionized state and only $\lesssim$10\% of the absorbing gas is neutral \citep{2011A&A...525A.113S}. In addition, there is no statistically significant correlation between the soft X-ray absorption and the dust column as inferred from optical/NIR data nor the metallicity of the neutral material \citep[e.g.,][]{2007MNRAS.377..273S, 2011arXiv1102.1469Z}, nor the darkness of a GRB \citep{2010MNRAS.402.2429C}. There is however a trend of higher visual extinctions with larger neutral metal columns, an anti-correlation between gas-to-dust ratio and metallicity \citep{2011arXiv1102.1469Z}, and dark bursts have stronger neutral metal absorption lines in their optical spectrum \citep{2011ApJ...727...73C}.

The dustiest afterglows and their hosts add two further hints. Firstly, there is the previously discussed anti-correlation between metals-to-dust ratio and sight-line extinction, albeit with a large scatter for individual events. And secondly, their hosts are on average redder, more luminous and massive, and supposedly also more evolved, dust- and metal-rich than their low-$A_V^{\rm GRB}$ counterparts \citep[e.g.,][]{2004ApJ...617..240K, 2005ApJ...635..260S}. There is hence a relation between the dust along the sight-line towards the GRB and physical host properties. Such a relation is expected if the dust probed by the afterglow is located at large enough distances to be fairly representative of the size of the host galaxy, its global dust enrichment and chemical state. In contrast, dust in the surrounding environment of a GRB, even if present and able to survive the intense afterglow or progenitor radiation, would rather be related to the very specific details of its circumburst environment or the GRB's birth cloud.

An overlap with optically selected hosts clearly exist: very young, blue and low-mass galaxies possess sight-lines through dusty regions as demonstrated through the afterglows and hosts of, e.g., GRBs~080605 and 100621A. For this kind of events the data are clearly not consistent with a uniform dust-shield. In a few cases the observations rather indicate a patchy dust distribution where the dust is located in clumps of small enough covering factor and large enough extinction to be negligible in the integrated host light distribution. The variations on the visual extinction in some cases are therefore a geometrical effect \citep[see also, e.g.,][]{2003ApJ...588...99B, 2009AJ....138.1690P}. 

This raises questions about the validity of star-formation estimates obtained for young and blue galaxies. Afterglow observations show that there is extinguished star-formation ongoing even in apparently unextinguished galaxies. Here, either the covering factor of the dusty clump is just too small to remove significantly from the host light, or the clump completely absorbs all UV light from the star-forming region, thus having a negligible effect on the host galaxy colors. In the latter case, the UV-derived SFRs would strictly represent lower limits on the ongoing star-formation in these galaxies. Far-infrared and sub-mm observations with $Herschel$ and ALMA, respectively, would enable to measure the galaxies SFR, dust mass and temperature and would help to clarify this issue.

A coherent picture of highly extinguished afterglows in combination with their diverse, but on average redder and more massive hosts could be obtained within a complex dust distribution made out of several constituents related to extinction in the diffuse ISM, extended interstellar clouds, or localized in fairly compact and dense regions such as giant molecular clouds.  

The dust extinction is hence very likely not directly associated with the GRB environment but plausibly with the neutral absorber at distances of a few hundred pc to one kpc. We stress, however, that the effective radii of dust destruction and photo-ionization will shape the detailed gas-to-dust and metals-to-dust ratios, adding further complexity to the absorbing system(s) in front of GRB afterglows. Furthermore, the dust distribution in high-$z$ galaxies could be even more complex due to dusty galaxy outflows, and reflect the absence of a uniform chemical enrichment on scales up to one kpc \citep[e.g.,][]{2009A&A...499...69N}. 


\section{Conclusions}


The afterglows of GRBs 081109 ($A_V^{\rm GRB} = 3.4_{-0.3}^{+0.4}\,\rm{mag}$) and 100621A ($A_V^{\rm GRB} = 3.8\pm0.2\,\rm{mag}$) join the growing sample of highly extinguished events. Their continuum emission is well-constrained by the combination of X-ray and NIR data, and the optical observations provide a detailed measurement of the dust properties along the sight-lines. While some diversity in their extinction properties, particularly dust abundance, clearly exists, GRBs 081109 and 100621A provide compelling evidence that a highly obscured afterglow is also a highly reddened one, and that extinction laws derived from local sight-lines accurately estimate the dust properties towards even highly extinguished GRBs.

Having a large enough sample of coeval afterglows with multi-wavelength data would ideally enable to advance from single sight-line, pencil beam investigations to a statistically symmetric geometry where each GRB afterglow represents a different sight-line through its host galaxy. In analogy to the case studied by \citet{1996ApJ...463..681W}, this could provide a good description of the structure and evolution of the absorbing medium and help constrain the opacity and filling factors of the dust geometry and clumps from the distribution of $A_V^{\rm GRB}$ values in star-forming galaxies out to very high redshift.

The hosts of the dustiest afterglows provide a different picture of GRB host galaxies compared to the hosts of optically-selected bursts. Although both samples are overlapping in their properties, the galaxies in this work have typical luminosities of around $L^{\ast}$ and stellar masses of $M_\ast \sim 10^{10} M_{\sun}$, more luminous and massive when compared to the hitherto discovered faint and blue hosts. Although the number counts are still low, this work indicates that a selection based on a large $A_V^{\rm GRB}$ picks up preferentially the more massive and chemically-evolved GRB hosts, which is in qualitative agreement with searches for dark GRB hosts \citep[][]{2010AAS...21540509P, 2011AAS...21710802P, Rossi2011}. 

This suggests that the properties of complete GRB host samples are diverse, and complex selection biases are still present: not only are the very faintest GRB hosts missing due to inherent sensitivity limits, but also some of the brightest, most luminous, and chemically evolved ones, because they have not been localized accurately enough. Fairly large and massive, dusty, and metal-rich galaxies are able to host GRBs, and the trend of low-metallicity GRB hosts is not as significant as claimed in previous studies, and possibly a selection effect of the young galaxy population dominating the global SFR at low-$z$ \citep[e.g.,][]{2007ApJ...660..504B, 2010arXiv1011.4506M}. This has substantial implications for the feasibility of tracing the star-formation history with GRB hosts and also for the progenitor channels of GRB production as a result of their metallicity dependence. In the former case, this work indeed indicates that the deficiency of high-mass GRB host galaxies in previous studies was at least partially due to a selection bias. The latter case, however, depends quite strongly on the assumption that the host-inferred metallicities are representative of the composition of the progenitor star, while different sight-lines through a GRB host can show a dispersion in metallicity of around a factor 100 \citep{2010MNRAS.402.1523P}.
  
Intriguingly, all GRBs with $A_V^{\rm{GRB}}\gtrsim4\,\rm{mag}$ have metals-to-dust ratios significantly below what is typically measured for GRB afterglows, and more in line with measurements from the Local Group. In addition, there is a strong anti-correlation between the metals-to-dust ratio and the visual extinction along the GRB sight-line. This effect seems independent on the specific host properties and can be interpreted as evidence of two physically independent absorbers:  dust-free, ionized metals in the circumburst environment (and probed by soft X-ray absorption), and in contrast a dusty absorber further out (and probed by reddening measurements in the UV/optical/NIR). 

A dust column independent of the immediate circumburst environment is further supported by the relation between afterglow $A_V^{\rm GRB}$ and host properties, in particular the on-average higher stellar mass and redder colors. Coupled with the blue and very young hosts of, e.g., GRBs 080605 or 100621A it indicates a complex dust geometry with different constituents in the diffuse ISM and in localized patches, which are plausibly associated with the cold-neutral absorber detected in rest-frame UV/optical spectra.

Further advances can now be made by getting direct observational access to more dusty sight-lines including $A_V^{\rm GRB} \gtrsim 10\,\rm{mag}$ events at an increasing redshift interval. Similar observations for a large enough sample would investigate the dependence of the global dust enrichment with cosmic evolution, and constrain the fraction of dust-enshrouded star-formation out to very high redshifts. A sophisticated observational strategy coupled with state-of-the-art instrumentation makes such a challenging study feasible: A rapid response of the order of several minutes by a NIR imager at an 8m-class telescope would have enabled the detection of the afterglows of GRB~100621A ($z\sim0.5$) up to $A_V^{\rm GRB} \approx 30\,\rm{mag}$, GRB~081109 ($z\sim1$) up to $A_V^{\rm GRB} \approx 20\,\rm{mag}$, and GRB~070802 ($z\sim2.5$) up to $A_V^{\rm GRB} \approx 10\,\rm{mag}$, which subsequently could have been followed-up using NIR spectroscopy. Once an accurate localization as well as detailed information about the GRB sight-line is available, their hosts should be readily accessible for multi-wavelength surveys via large ground- and space-based facilities, yielding information about otherwise fully extinguished environments and unprecedented insights into the conditions of star-forming galaxies throughout the Universe.
  
\begin{acknowledgements}
We thank the referee for valuable comments. TK acknowledges support by the DFG cluster of excellence 'Origin and Structure of the Universe' and support by the European Commission under the Marie Curie IEF Programme in FP7. Part of the funding for GROND (both hardware as well as personnel) was generously granted from the Leibniz-Prize to Prof. G. Hasinger (DFG grant HA 1850/28-1). The Dark Cosmology Centre is funded by the Danish National Research Foundation. PS acknowledges support by DFG grant SA 2001/1-1. SS acknowledges support through project M.FE.A.Ext 00003 of the MPG. SK, DAK \& ANG acknowledge support by DFG grant Kl 766/16-1. ARo acknowledges support from the BLANCEFLOR Boncompagni-Ludovisi, n\'ee Bildt foundation. SMB acknowledges support of a European Union Marie Curie European Reintegration Grant within the 7th Program under contract number PERG04-GA-2008-239176. This work made use of data supplied by the UK Swift Science Data Centre at the University of Leicester.

\end{acknowledgements}
\hyphenation{Post-Script Sprin-ger}


\end{document}